\RequirePackage{fix-cm}
\documentclass[smallextended]{svjour3}       
\smartqed  
\usepackage{amsmath}
\usepackage{amssymb}
\usepackage{latexsym}
\usepackage{graphicx}
\begin{document}

\title{Thermodynamics and Geometrothermodynamics of Charged black holes in Massive Gravity
}


\author{Jishnu Suresh         \and C P Masroor \and Geethu Prabhakar \and
        V C  Kuriakose
}


\institute{Jishnu Suresh. \at
              Cochin University of Science and Technology \\
              Cochin-682022\\
              \email{jishnusuresh@cusat.ac.in}      
           \and
            C P Masroor\at
              School of Pure and Applied Physics, Mahatma Gandhi University, Kottyam, India\\
              \email{masroorcp@gmail.com}     
           \and
        Geethu Prabhakar\at
               School of Pure and Applied Physics, Mahatma Gandhi University, Kottyam, India\\
              \email{geethuprabhakar3@gmail.com}     
            \and
           V C  Kuriakose. \at
              Cochin University of Science and Technology\\
              Cochin-682022\\
              \email{vck@cusat.ac.in}
}

\date{Received: date / Accepted: date}

\maketitle

\begin{abstract}
The objective of this paper is to study the thermodynamics and thermodynamic
geometry of charged de-Sitter and charged anti de-Sitter black hole solutions in massive
gravity. We study the effect of curvature parameter as well as the mass of graviton in the
thermodynamics of the black hole system. We further extend our studies to different
topology of the space time and its effects on phase transition and thermodynamics. In
addition, the phase transition structure of the black hole and its interactions are
reproduced using geometrothermodynamics.
\end{abstract}

\section{Introduction}
It is being believed that the idea of incorporating a small mass to the graviton will
definitely lead to a new modified theory of gravity. 
The first attempt towards the realization of this theory was initiated by Fierz and Pauli \cite{fierz}.
They added a quadratic mass to the action as,
$$\mathcal{L} = - \frac{1}{8} m^2 (h_{\mu \nu} ^2 - A h^2).$$ But it is
observed that this linearized Fierz-Pauli theory violates
the gauge invariance in general relativity. Considering the linear Fierz-Pauli 
theory in the zero limit of graviton mass, i.e., $m\rightarrow 0$, 
the theory was unable to recover the general relativity. This mismatch leads to 
contradiction with the solar system tests due to 
van Dam-Veltman-Zakharov discontinuity \cite{vdvz}, regardless of smallness of graviton mass, 
the relativistic and non relativistic matter 
couple to gravity with different relative strengths. By introducing the nonlinear effects
using Vainshtein \cite{vainshtein} mechanism, the problems due to 
vDVZ discontinuity can be solved out. But the massive gravity theory in this nonlinear situation, losses 
the momentum constraints as well as
the Hamiltonian constraints and they lead to a situation where the gravity suffers from an instability 
known as the Boulware-Deser (BD)ghost \cite{boulware}, as a
result of an extra degree of freedom other than the five degrees of freedom of a massive
spin-2 field in a Poincare invariant
background. Recently de Rham-Gabadadze-Tolley (dRGT) \cite{dRGT1,dRGT2} has introduced a theory which is free from
the BD ghost. The dGRT theory consists of a three 
parameter family of potentials, which in turn produce a dynamical constraint which
can be used to remove the degree of freedom which acts as the BD 
ghost. Even the absence of BD ghost does not make the theory consistent in every manner,
the diffeomorphism invariance broken due to mass term
was one of the major problem. The St$\ddot{\mbox{u}}$ckelberg mechanism \cite{arkani} can be used to overcome 
this situation. It can be done by introducing four auxiliary 
scalars $\phi^A (x)$ via $\bar g_{\mu \nu} \rightarrow \partial_{\mu} \phi^A \partial_{\nu} \phi^B \eta_{AB} $. 
Using this pure gauge 
St$\ddot{\mbox{u}}$ckelberg field, the original theory, 
the general relativity can be recovered using 
a unitary gauge, $\phi^A = \delta_{\mu} ^{A} x^{\mu}.$ Hence this nonlinear St$\ddot{\mbox{u}}$ckelberg formalism brings back the diffeomorphism invariance.  Recently, interesting black hole solutions of massive gravity have been found in \cite{sol1,sol2,sol3,sol4,sol5}.

Various black hole and cosmological solutions of the massive gravity theory are intensively studied in recent years. 
Black holes receive  attention of the many, as they are identified as thermodynamic objects with a physical 
temperature and entropy \cite{bekenstein,hawking1,hawking2} as well as there exist an analogy
between black hole system and the non gravitational thermodynamic system \cite{bardeen}. Hence exploring the thermodynamics
of black holes will eventually give many ideas regarding the corresponding 
gravity theory of the black hole system. 
The analysis of black hole phase transitions, which are entropically driven in Einstein's theory, 
but depends on other parameters in modified 
theories are of prime importance in black hole thermodynamic studies. Black hole phase transition studies can be done in two ways. 
One, by calculating the heat capacity of black hole space time and there by studying the transition between thermodynamically
stable phase and unstable phase. 
The second method is by incorporating differential geometry ideas in to black hole thermodynamics.
Recently, this geometric method gained attention, since it can be used as a convenient tool to study 
the thermodynamics and the corresponding phase transition structures of black holes. 
Various investigations are done by incorporating the idea of contact geometry in to the study of 
black hole thermodynamics \cite{gibbs,caratheodory,hermann,mrugala1,mrugala2}. 
The recent method in this context is a new geometric formalism, known as the Geometrothermodynamics 
proposed by Quevedo et.al \cite{quevedo1,quevedo2,quevedo3}. This method can give exact explanations 
for all the  thermodynamic behaviors as well as the exact phase structure of black hole systems. 

A brief outline of the paper is as follows: in Section 2, we review the ideas of geometrothermodynamics. 
We explain the charged black hole solutions in massive gravity and calculate the relevant thermodynamic quantities like the horizon
temperature, entropy and the specific heat in Section 3. Geometrothermodynamic method for analyzing 
the phase transition picture are depicted in Section 4. The results are summarized in Section 5

\section{Black Holes in Massive Gravity}
The massive gravity model which is used in this paper can be described using the action \cite{dRGT2},
\begin{equation}
 S=\int d^D x \left[ \frac{M^2_{pl}}{2} \sqrt{-g} \left(R + m^2 \textit{U}(g,H)\right) \right],
\end{equation}
where the first term is the usual Einstein-Hilbert action and the second term is arising from the contributions of mass of the 
graviton $m$, and from the nonlinear higher derivative term $U$ corresponding to the massive graviton. It is given by
\begin{equation}
 U=U_2 + \alpha_3 U_3 + \alpha_4 U_4 ,
 \label{ghost_free_potential}
\end{equation}
where,
\begin{eqnarray}
 U_2 &=& [\mathcal{K}]^2 - [\mathcal{K} ^2] \nonumber \\
 U_3 &=& [\mathcal{K}]^3 - [\mathcal{K}] [\mathcal{K}^2] + 2 [\mathcal{K}^3] \nonumber \\
 U_4 &=& [\mathcal{K}]^4 - 6 [\mathcal{K}]^2 [\mathcal{K}^2] + 8[\mathcal{K}^3][\mathcal{K}] - 6[\mathcal{K}^4]. 
\end{eqnarray}
In the above set of equations, the tensor $\mathcal{K}_{\nu} ^{\mu}$ is defined as,
\begin{equation}
\mathcal{K}_{\nu} ^{\mu} = \delta_{\nu} ^{\mu}-\sqrt{\partial^{\mu} \phi^{\alpha} \partial_{\nu} \phi^{\beta} f_{\alpha \beta}} ,
\end{equation}
where $\phi^{\alpha}$ and $\phi^{\beta}$ are the corresponding Stu$\ddot{c}$kelberg field and $f_{\alpha \beta}$ is a fixed symmetric
tensor usually called as the reference metric.

In the unitary gauge, defined 
as $ \phi^a = x^a $, the term $h_{\mu\nu}\,=\,g_{\mu\nu}\,-\,\eta_{\mu\nu}$ is the gravitational analog of the Proca field of 
massive electrodynamics \cite{Berezhiani}. By introducing the St\"{u}ckelberg field $\phi^a$, which can be considered as background field plus a pion contribution,
$\phi^a \,=\, x^a \,+ \, \pi^a$\cite{koyama}, and replacing the Minkowski metric by,
$$g_{\mu\nu} = \partial_\mu \phi^a \partial_\nu \phi^b \eta_{ab}+H_{\mu\nu},$$
where $H_{\mu\nu}$ is the covariantized metric perturbation, 
one can restore the Diffeomorphism invariance.
As given in \cite{koyama,Berezhiani}, two new coefficients $ \alpha  $ and $\beta$  
are introduced which relate
the coefficients $\alpha_3$ and $\alpha_4$ in (\ref{ghost_free_potential}) by, 
\begin{equation}
\alpha_3 = -\frac{(-\alpha+1)}{3},
\end{equation}
 and 
\begin{equation} 
\alpha_4=\frac{-\beta}{2}+\frac{(-\alpha+1)}{12}.
\end{equation}
In empty space, the equation of motion is given as,
\begin{equation}\label{eq:source_free_massive_eom}
G_{\mu\nu} + m^2 X_{\mu\nu} = 0,
\end{equation}
where $ X_{\mu\nu} $ is the effective energy-momentum tensor contributed by the graviton mass $ m $, which is given by,
\begin{eqnarray}
X_{\mu\nu} & = & -\frac{1}{2} \left[ K g_{\mu\nu} - K_{\mu\nu} + \alpha \left(K^2_{\mu\nu} - K K_{\mu\nu} +
\frac{1}{2} g_{\mu\nu}([K]^2 - [K^2])\right) \right. \nonumber \\
&   & + 6 \beta \left( K^3_{\mu\nu} - K K^2_{\mu\nu} + \frac{1}{2} K_{\mu\nu}([K]^2-[K^2]) \right. - \frac{1}{6} g_{\mu\nu} ([K]^3 \left. \left. -3[K][K^2] + 2[K^3])\right) \right].
\label{energy_momentum tensor}
\end{eqnarray}
Now applying the Bianchi identity, $ \nabla^\mu G_{\mu\nu} = 0 $ in (\ref{eq:source_free_massive_eom}), we arrive at the constraint equation,
\begin{equation}
m^2\nabla^\mu X_{\mu\nu} = 0. 
\label{constraint}
\end{equation}
The parameters of the action, namely $\alpha$ and $\beta$ can be chosen in different ways so that one ends up with different black hole solutions. Particularly for the choice $\beta=\alpha^2$, the space of solution is much wider than the general case discussed in \cite{sol3}. As a result, this choice lead to much richer family of solutions compared to the general choice of $\alpha$ and $\beta$. Here, we concentrate on a particular family of the ghost-free theory of massive gravity \cite{koyama,Berezhiani,Nieuwenhuizen}, where,
\begin{equation}
\beta = -\frac{\alpha^2}{6}.
\end{equation}
For this special choice, introduced in \cite{Nieuwenhuizen}, (\ref{constraint}) is automatically satisfied for a certain diagonal and time-independent metrics in  spherical polar coordinates. 
One can consider this as a limitation, that these exact analytic black hole solutions are obtained only for a specific choice of the two free parameters of massive gravity. But such a choice of parameters is peculiar because on this background the kinetic terms for both the vector and scalar fluctuations vanish in the decoupling limit. Hence one would expect infinitely strong interactions for these modes. To account for this issue, we would consider these solutions bellow as just an example for demonstrating how non-singular solutions could emerge as well as how their thermodynamic properties and interactions behave in the presence and absence of the massive parameters. 

Now using (\ref{eq:source_free_massive_eom}), (\ref{energy_momentum tensor}) and (\ref{constraint}), a spherically symmetric and time independent metric in de Sitter space
can be obtained, by choosing,
\begin{equation}
m^2X_{\mu\nu} = \lambda g_{\mu\nu},
\label{condition_ds}
\end{equation}
where $ \lambda $ is a constant.
The solutions of (\ref{eq:source_free_massive_eom}) that satisfies the condition given in (\ref{condition_ds}) with a 
positive but otherwise arbitrary $\alpha$ is given by,
\begin{equation}
 ds^2 = -\kappa^2 dt^2 + \left( \frac{\alpha}{\alpha +1} dr \pm \kappa \sqrt{\frac{2}{3 \alpha}} \frac{\alpha}{\alpha +1} m r dt \right) ^2 
         + \frac{\alpha^2}{(\alpha +1)^2} r^2 d \Omega ^2 .
\end{equation}
Here $\kappa$ is a positive integration constant.  The obtained solution is free of singularities. 
 Now by coupling this ghost-free massive gravity theory to
the Maxwell's theory of electromagnetism, one can obtain the Reissner- Nordstr\"{o}m solution in dS space as,
 \begin{equation}
ds^2 = -dt^2 + \left( \tilde{\alpha}dr \pm \sqrt{\frac{r_g}{\tilde{\alpha}}
+ \frac{2 \tilde{\alpha}^2}{3 \alpha} m^2r^2 - \frac{\tilde{Q}^2}{\tilde{\alpha}^4r^2}dt} \right)^2 + \tilde{\alpha}^2r^2d\Omega^2,
\label{MRNds_original}
\end{equation} 
here $ \tilde{\alpha} \equiv \alpha/(\alpha +1) $, $m$ is mass of graviton, $\alpha$ is the curvature parameter and the 
electromagnetic field is given by,
\begin{equation}
E = \frac{\tilde{Q}}{r^2}\;\;\;\;\;\;\; \mbox{and} \;\;\;\;\;\;\; B =0.
\end{equation}
To rewrite the above charged dS solution in arbitrary space time in the static slicing,
one can make the following transformations for spatial and temporal coordinates respectively as,
\begin{equation}
r \rightarrow \frac{r}{\tilde{\alpha}},
\end{equation}
and,
\begin{equation}
dt \rightarrow dt+f'(r)dr,
\end{equation}
where,
\begin{equation}
f'(r) \equiv -\frac{g_{01}}{g_{00}} = \pm \frac{\sqrt{\frac{r_g}{r} + \frac{2}{3\alpha}m^2 r^2 
+ \frac{\tilde{Q}^2}{\tilde{\alpha}^2 r^2}}} {k-\frac{r_g}{r} - \frac{2}{3 \alpha} m^2 r^2 + \frac{\tilde{Q}^2}{\tilde{\alpha}^2 r^2}}.
\end{equation}
Once these transformations are performed, one can easily rewrite (\ref{MRNds_original}) in a familiar form as,
\begin{equation}
ds^2 = -\left(k - \frac{r_g}{r} - \frac{2}{3\alpha}m^2r^2 
+ \frac{\tilde{Q}^2}{\tilde{\alpha}^2 r^2} \right) dt^2 + \frac{dr^2}{k - \frac{r_g}{r} - \frac{2}{3\alpha}m^2r^2 
+ \frac{\tilde{Q}^2}{\tilde{\alpha}^2 r^2}} + r^2 d\Omega^2.
\label{MRNds_transformed}
\end{equation}
Due to the above transformations, the St\"{u}ckelberg field becomes,
\begin{equation}
 \phi^0 = t + f(r),
\end{equation}
\begin{equation}
 \phi^r = r+ \frac{1}{\alpha} r ,
\end{equation}
and electromagnetic field is, 
\begin{equation}
 E = \frac{\tilde{Q}}{\tilde{\alpha}r^2}\;\;\;\;\;\;\; \mbox{and} \;\;\;\;\;\;\; B =0.
\end{equation}
where the actual charge should be redefined as $Q \equiv \tilde{Q}/ \tilde{\alpha}$. 

Using (\ref{MRNds_transformed}) and solving for  $f(r)=0$ at the horizon 
limit and incorporating the Bekenstein-Hawking area law, $S=\frac{A}{4}$,
one can easily obtain the mass of the black hole as,
\begin{equation}
 M=  \frac{3 \pi^2 \tilde{Q}^2 \alpha + \tilde{\alpha} S \left( 3k \pi \alpha - 2 m^2 S \right)}
     {6 \tilde{\alpha}^2 \pi^{3/2} \sqrt{S} \alpha}.
     \label{mass}
\end{equation}
From the first law of thermodynamics, $\delta M=T \delta S + \Phi \delta Q$,
temperature $T$ can be calculated as,
\begin{equation}
 T= \frac{\tilde{\alpha} S \left( k \pi \alpha -2 m^2 S \right) - \pi^2 \tilde{Q}^2 \alpha}
    {4 \tilde{\alpha}^2 \pi^{3/2} S^{3/2} \alpha}.
    \label{temperature}
\end{equation}
Using the classical thermodynamic relation $C=T\left(\frac{\partial S}{\partial T}\right)$, one can obtain the 
heat capacity of the black hole as,
\begin{equation}
 C= \frac{2S \left( \pi^2 \tilde{Q}^2 \alpha + \tilde{\alpha} S (2m^2 S -k \pi \alpha) \right)}
    {\tilde{\alpha}^2 S (k \pi \alpha + 2 m^2 S) - 3 \pi^2 \tilde{Q}^2 \alpha}.
    \label{heatcapacity}
\end{equation}

It is interesting to note that, the charged black hole solution in massive gravity (\ref{MRNds_transformed}) with $m$ as
the mass of the graviton can act as Reissner- Nordstr\"{o}m solution in de-Sitter and anti de-Sitter space time in massive gravity with respect
to the choice of curvature parameter $\alpha$. We can see that (\ref{MRNds_transformed}) will  reduce to RNdS solution in massive gravity for $\alpha>0$, 
for $\alpha<0$ we can obtain RNAdS solution and the RN solution in Einstein's general relativity for mass of the graviton $m=0$.

Before going in to the details of phase transition structure of charged black holes in massive gravity, let us consider
the physics of the curvature parameter $\alpha$ in details. From the black hole space time metric (\ref{MRNds_transformed}),
and comparing it with the charged dS or AdS black hole in Einstein's gravity theory, one can easily 
identify the cosmological constant term in massive gravity as, 
\begin{equation}
 \Lambda=\frac{m^2}{\alpha}.
 \label{cosmologicalconstant}
\end{equation}
We know that in most black hole thermodynamic studies, the cosmological constant is treated as a fixed parameter. But it
has been suggested that it is better to treat $\Lambda$ as a thermodynamic variable \cite{HT,Teitelboim}. In many studies, this cosmological
constant is treated as the thermodynamic variable, the pressure \cite{dolan1,dolan2}. Accordingly the cosmological constant generated pressure can be
written as, 
\begin{equation}
 P=-\frac{\Lambda}{8 \pi G}.
\end{equation}
 From (\ref{cosmologicalconstant}), one can rewrite the pressure generated by cosmological constant in massive gravity as,
 \begin{equation}
  P=-\frac{m^2}{8\pi \alpha}.
 \end{equation}
Hence the choice of curvature parameter $\alpha$ will determine whether the space time is de Sitter or anti de Sitter. 
The presence of negative pressure, by choosing the curvature parameter as positive, point towards the accelerated expansion of the present
universe. Further studies in this direction may lead to a better understanding of this phenomenon. 
Throughout this paper we consider the dependency of curvature parameter on the thermodynamic behaviors hence indirectly the effects of 
cosmological constant and pressure.  
\begin{figure}[h]
              \includegraphics[scale=0.4]{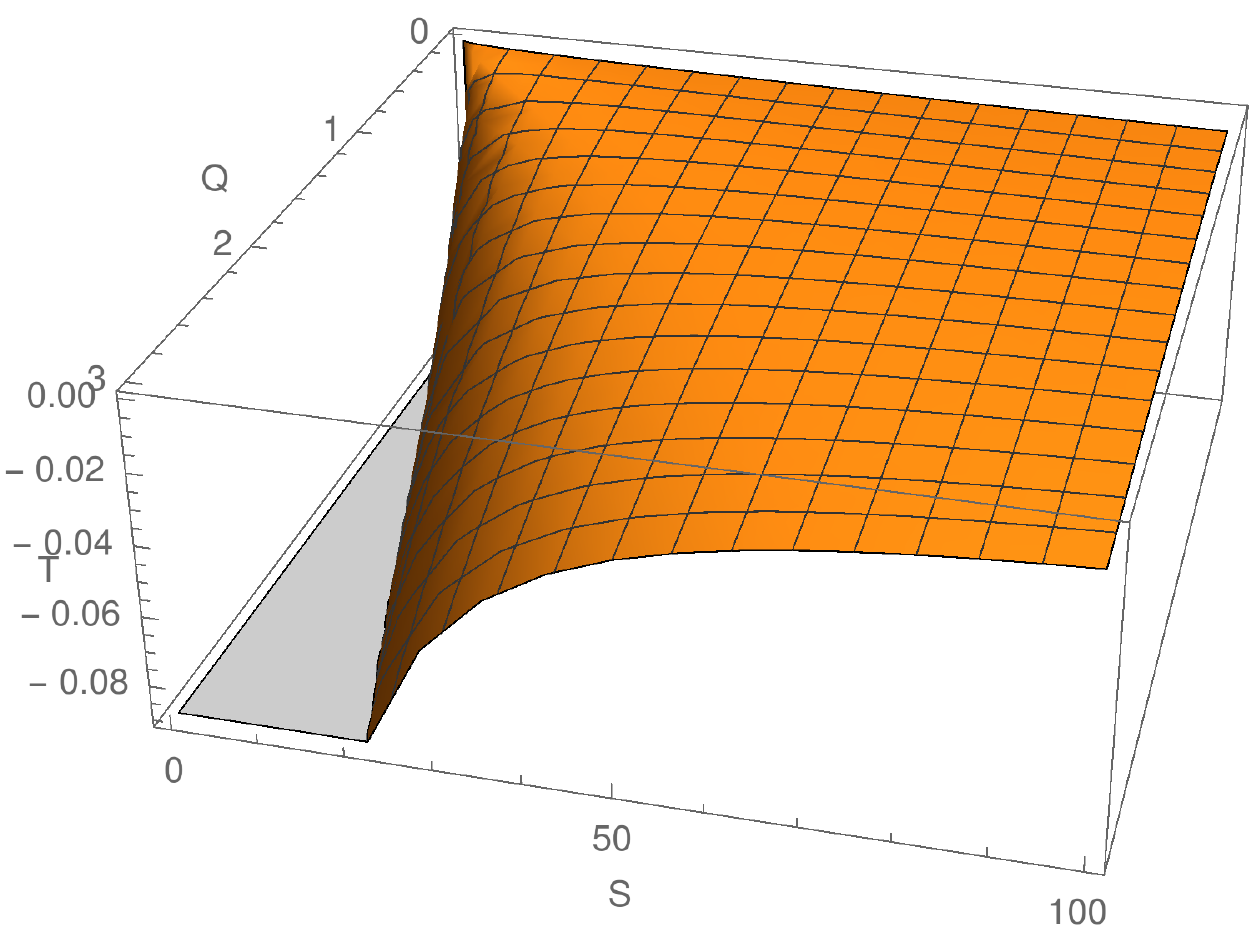}
	      \hspace{1cm}
              \includegraphics[scale=0.4]{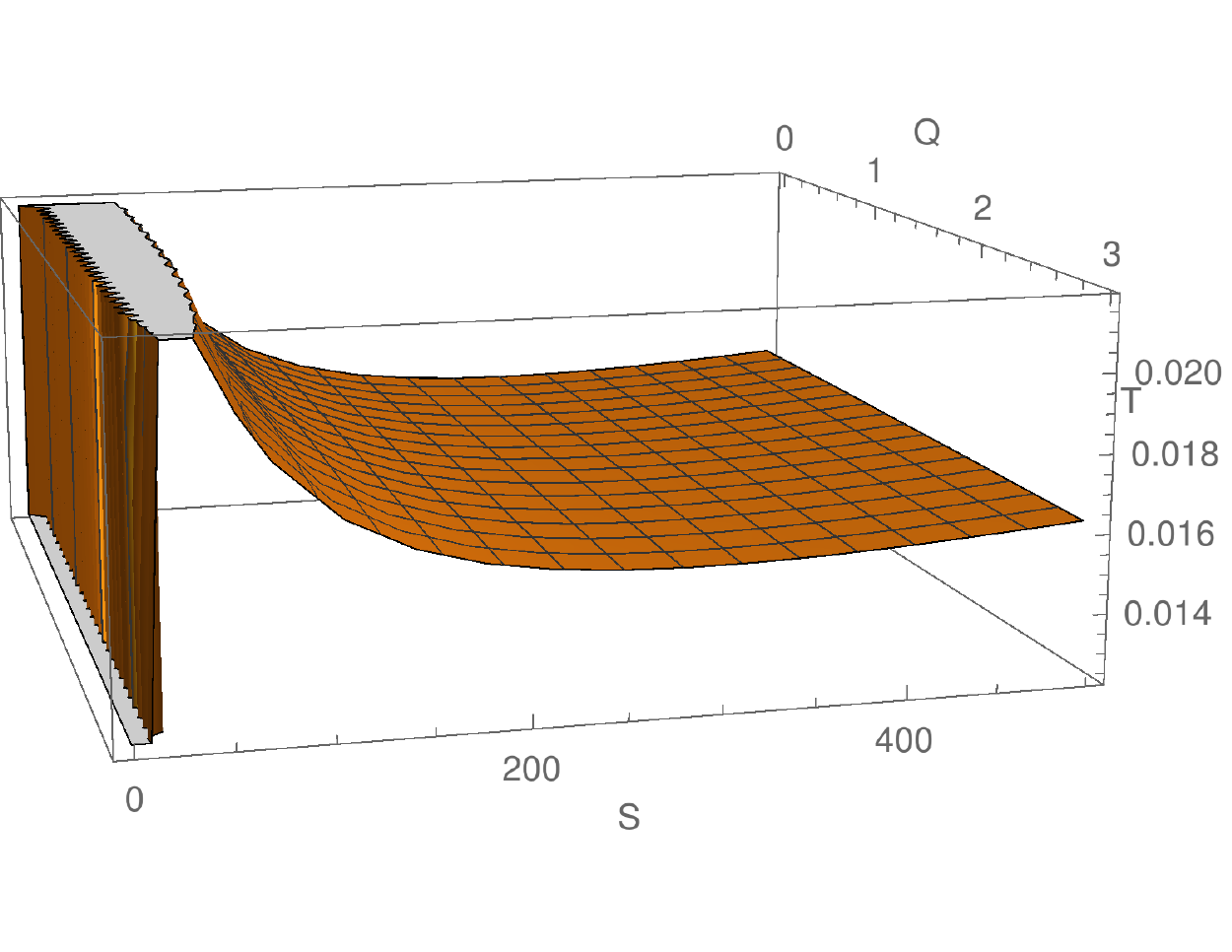}
       	      \hspace{1cm}
              \includegraphics[scale=0.4]{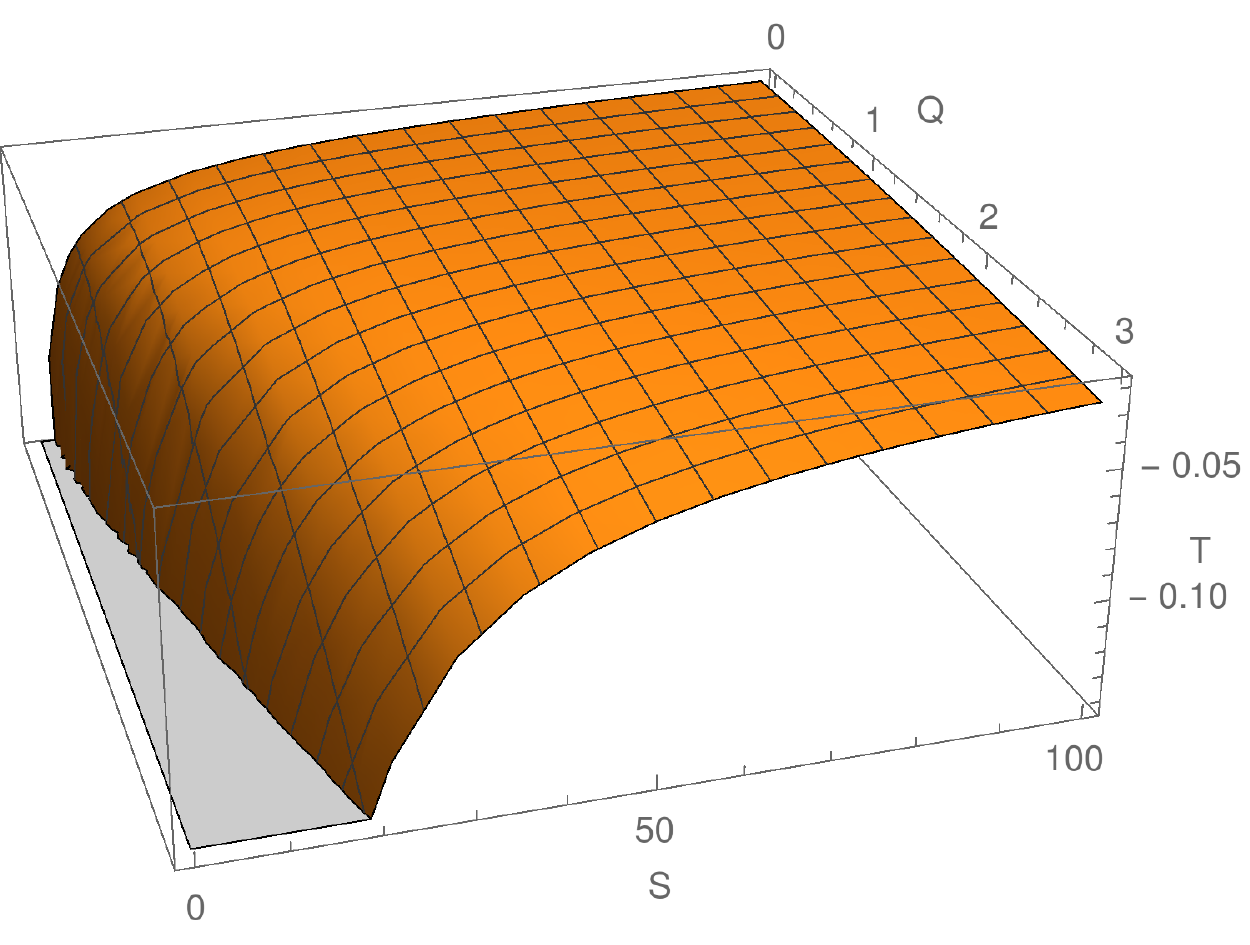}
              \caption{3D Variation of temperature against entropy and charge for de-Sitter black holes for flat, spherical and hyperbolic topology of space time in massive gravity}
  \label{temp_ds_fig} 
 \end{figure}
\subsection{Charged de-Sitter black hole in massive gravity}
\begin{figure}
              \includegraphics[scale=0.4]{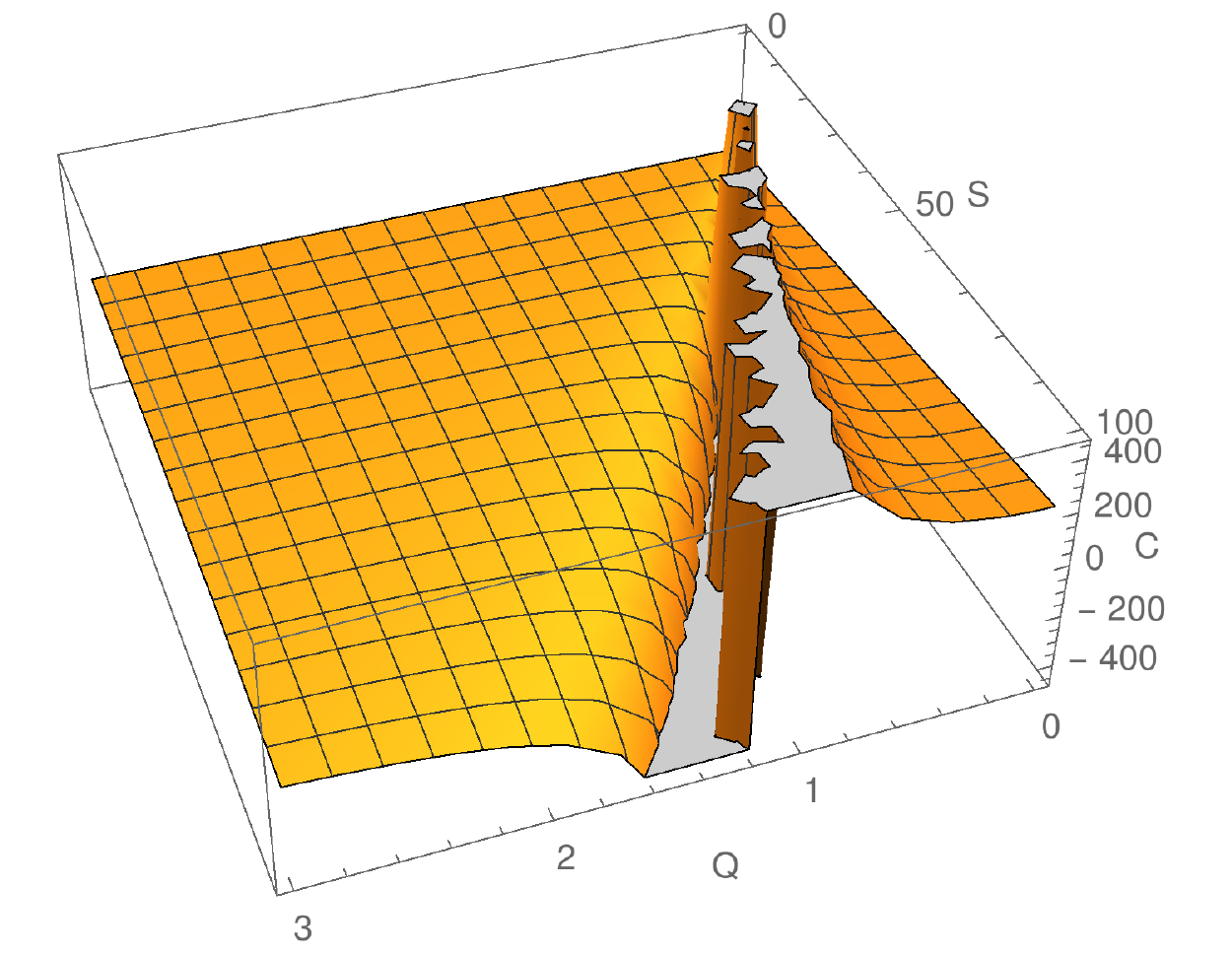}
	      \hspace{1cm}
              \includegraphics[scale=0.4]{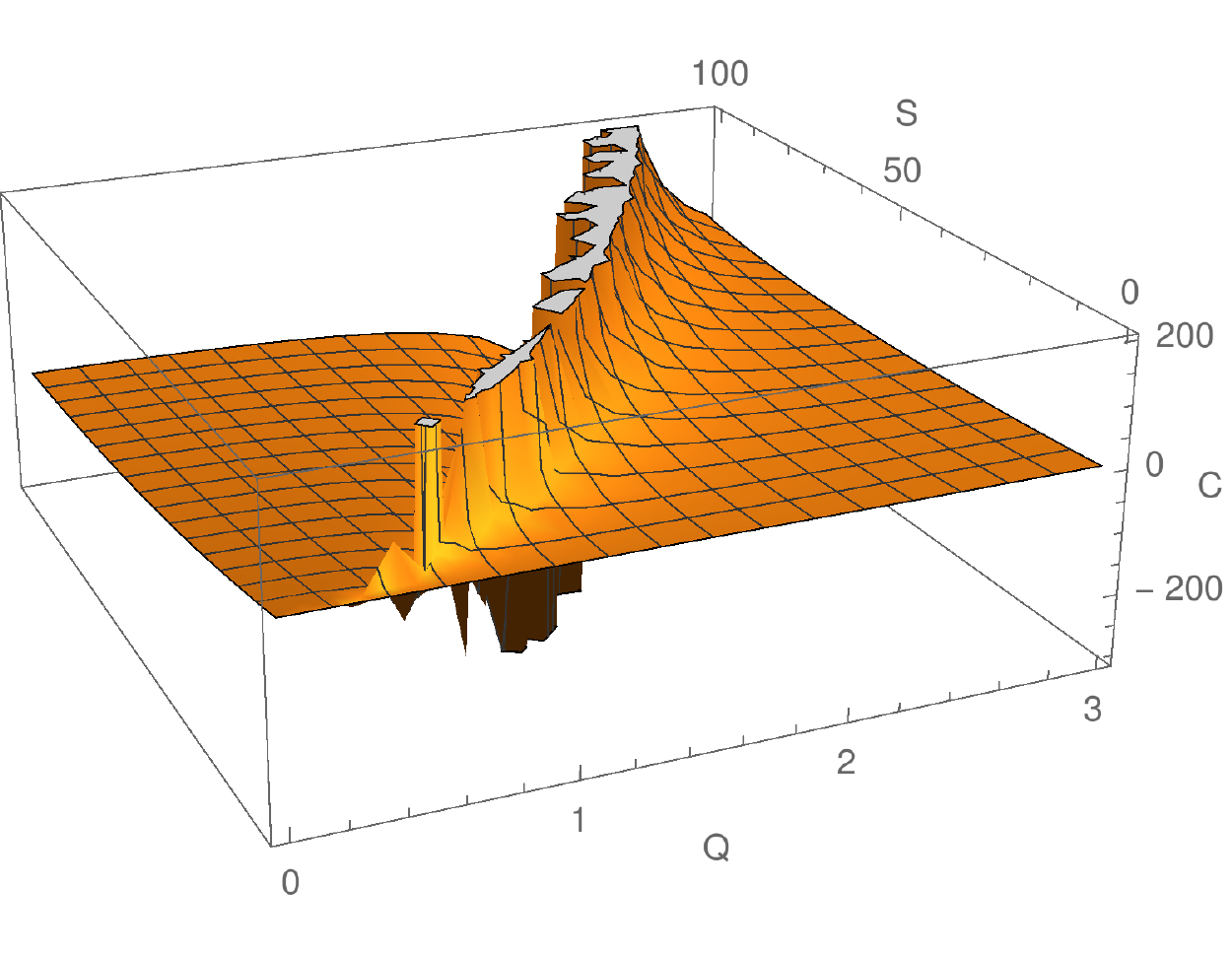}
       	      \hspace{1cm}
              \includegraphics[scale=0.4]{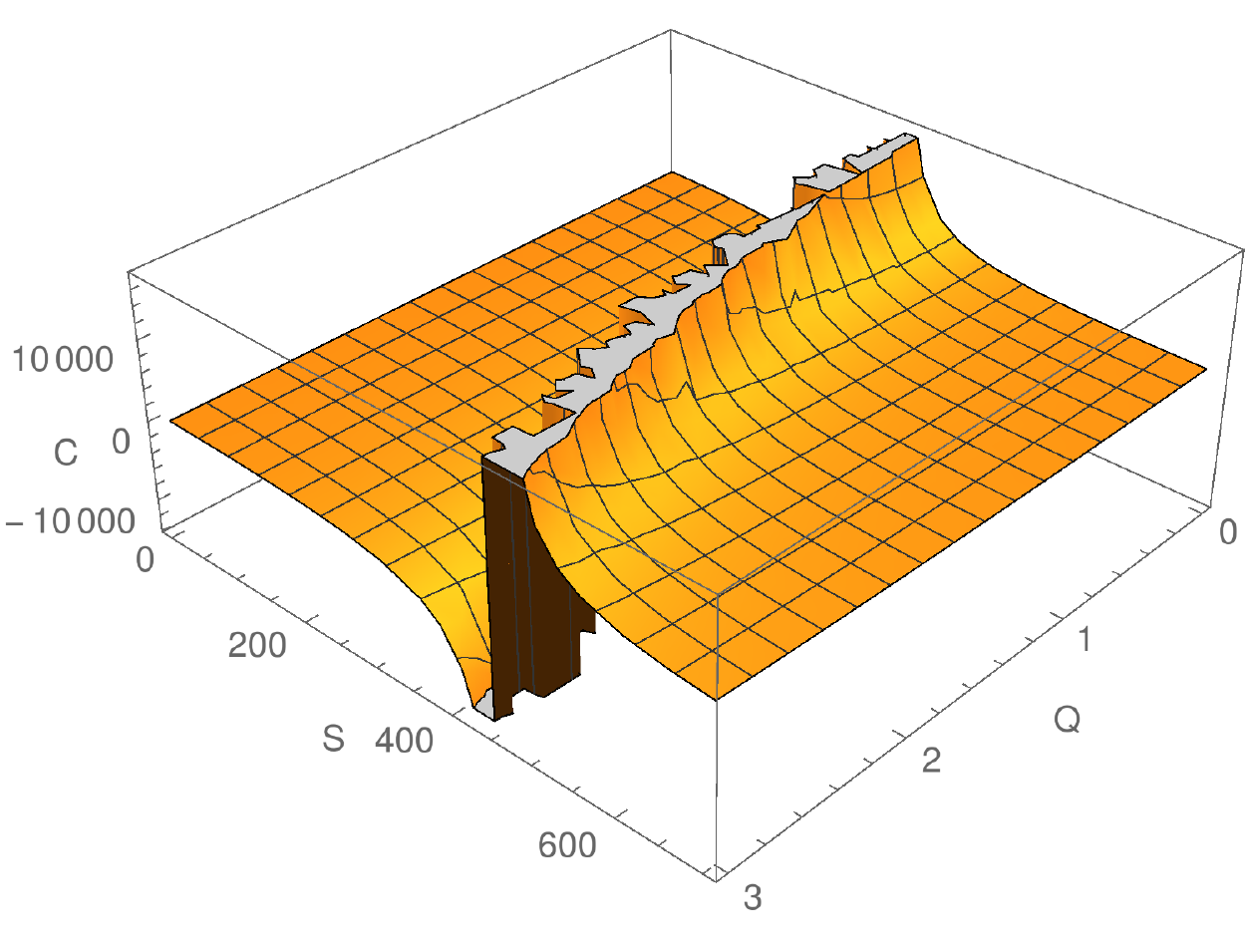}
              \caption{3D Variation of heat capacity against entropy and charge for de-Sitter black holes for flat, spherical and hyperbolic topology of space time in massive gravity}
              \label{spec_ds_fig}  
 \end{figure}

Let us consider the case where the curvature parameter is taken to be
positive ($\alpha>0$), then the black hole solution given by (\ref{MRNds_transformed}) reduces to charged
de-Sitter solution in massive gravity (RNdS). Now the variation of temperature and specific heat with 
entropy of RNdS black hole is plotted for different space time in figures (\ref{temp_ds_fig}) and (\ref{spec_ds_fig}).

From figure (\ref{temp_ds_fig}) we can see that for the flat case ($k=0$), the temperature is always negative, and hence has no physical significance.
For the spherically symmetric space time case ($k=1$), the temperature initially enters in to a physically insignificant 
region (with negative temperature) and lies in a positive region for intermediate sized black holes. For black holes with larger
horizon radius it again goes to the negative temperature regions. For the hyperbolic space time case ($k=-1$), the behaviour exactly
resembles that of flat space time. So for the RNdS case in massive gravity for spherically symmetric space time, there exists a 
window at which the black hole has positive temperatures and hence lies in a physically significant region.

The variation of heat capacity depicted in figure (\ref{spec_ds_fig}) shows that, black hole undergoes phase transitions for all
the cases, $k=0,1,-1$. For flat and hyperbolic space time cases, the black hole lies in a thermodynamically unstable phase
and undergoes a infinite discontinuity and become
thermodynamically stable. But for spherically symmetric space time, the black hole goes from thermodynamically 
stable  region to an unstable region at many points including the infinite discontinuity transitions for black hole with lower horizon radius.
 \begin{figure}
              \includegraphics[scale=0.4]{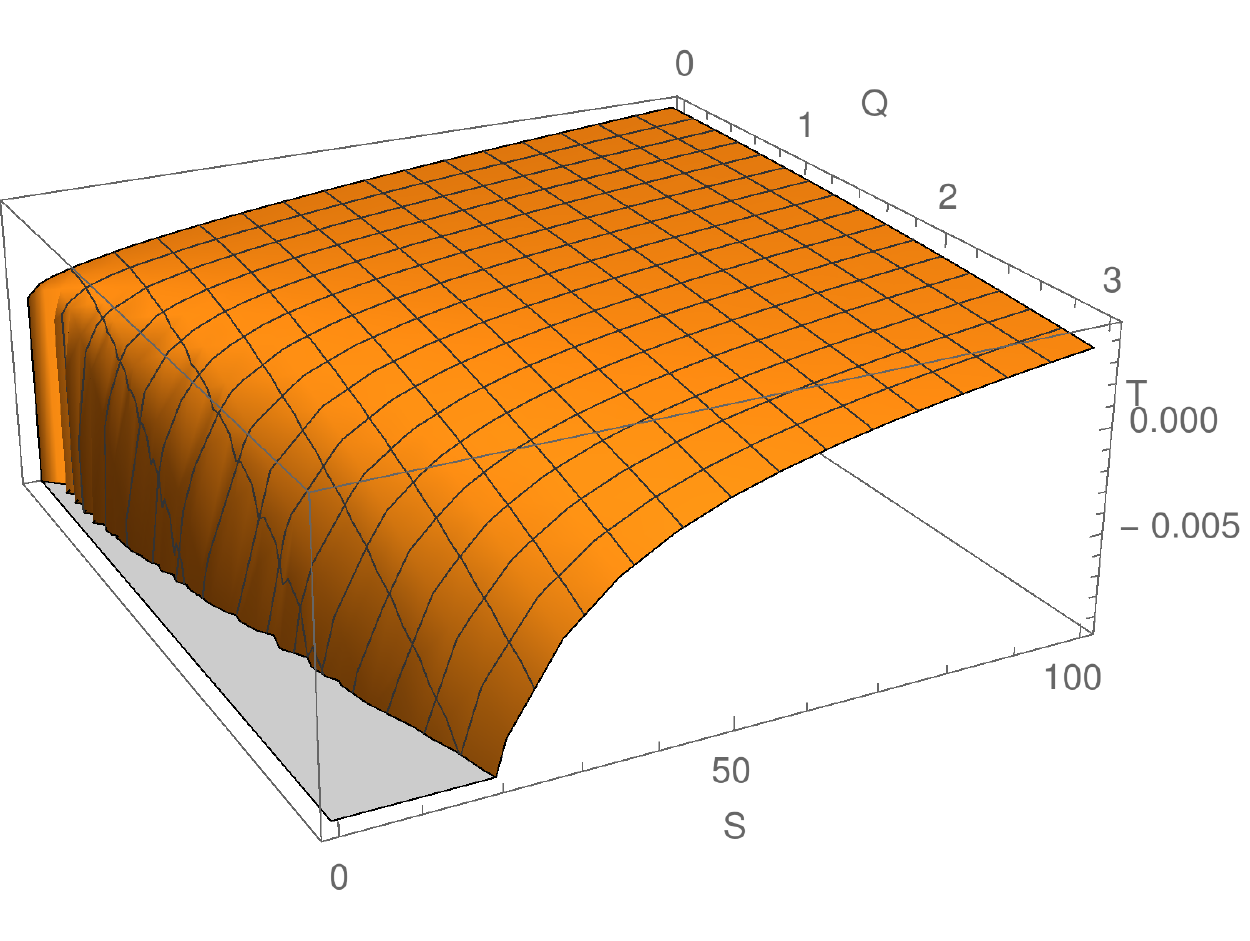}
	      \hspace{1cm}
              \includegraphics[scale=0.4]{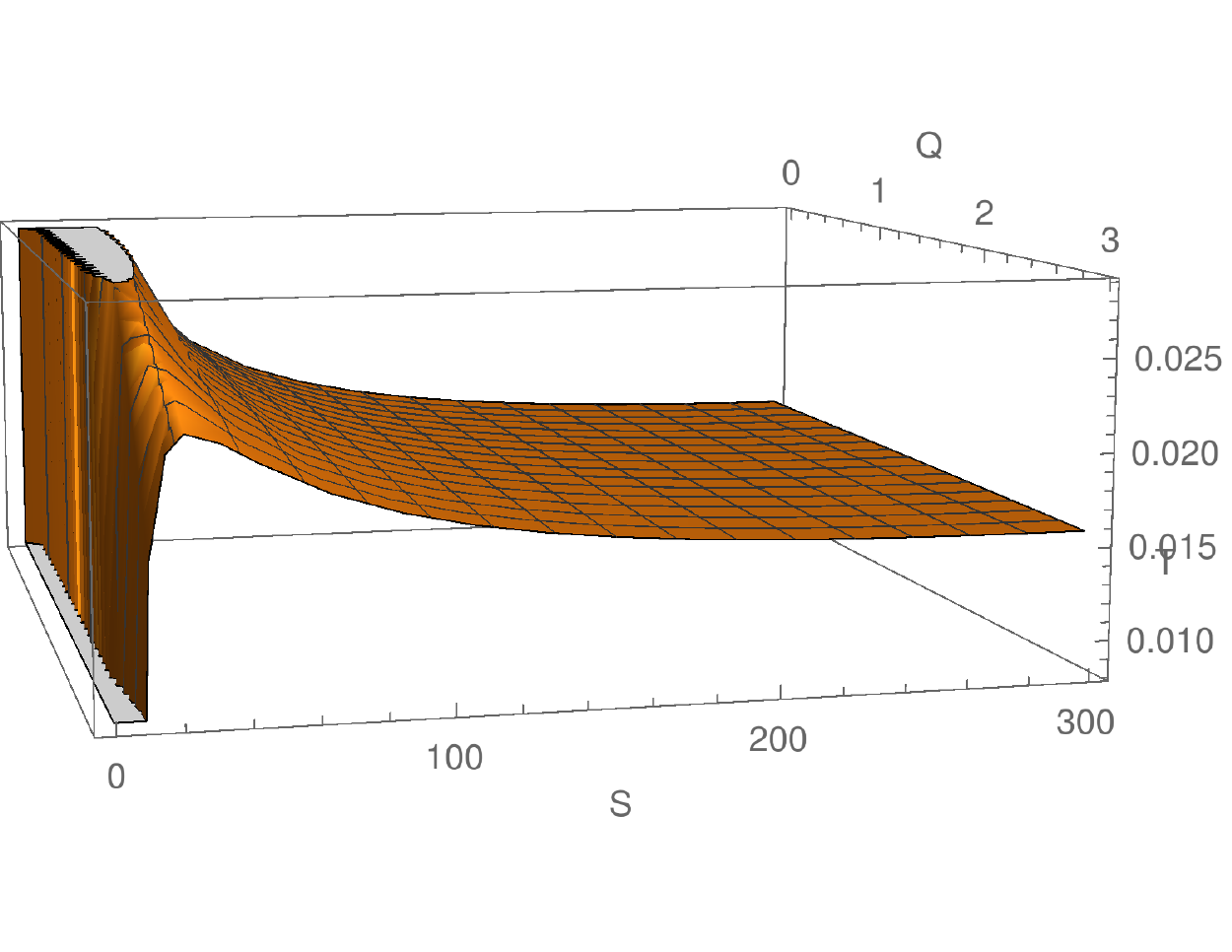}
       	      \hspace{1cm}
              \includegraphics[scale=0.4]{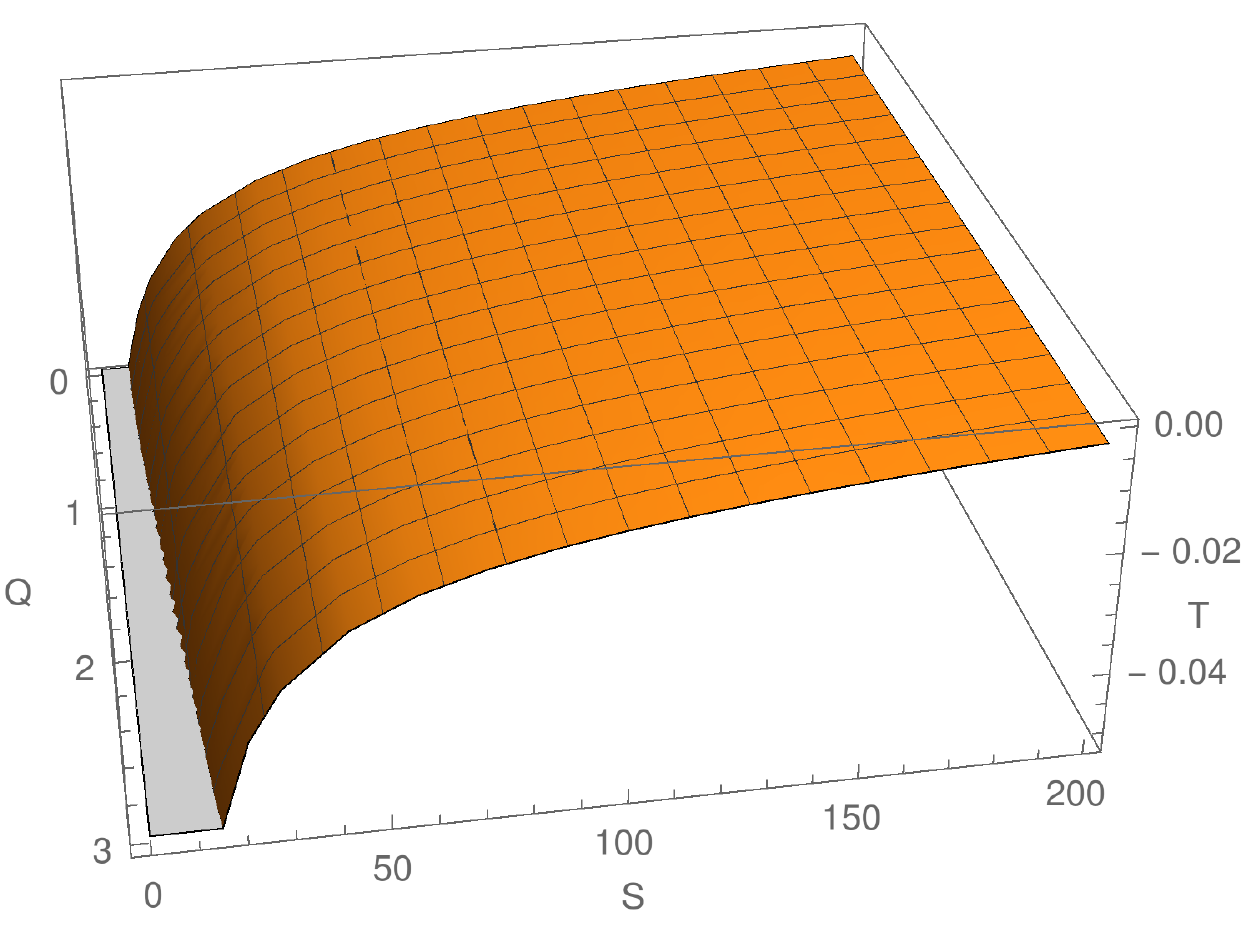}
              \caption{3D Variation of temperature against entropy and charge for anti de-Sitter black holes for flat, spherical and hyperbolic topology of space time in massive gravity}
               \label{temp_ads_fig}  
 \end{figure}
\subsection{Charged anti de-Sitter black hole in massive gravity}
\begin{figure}
              \includegraphics[scale=0.4]{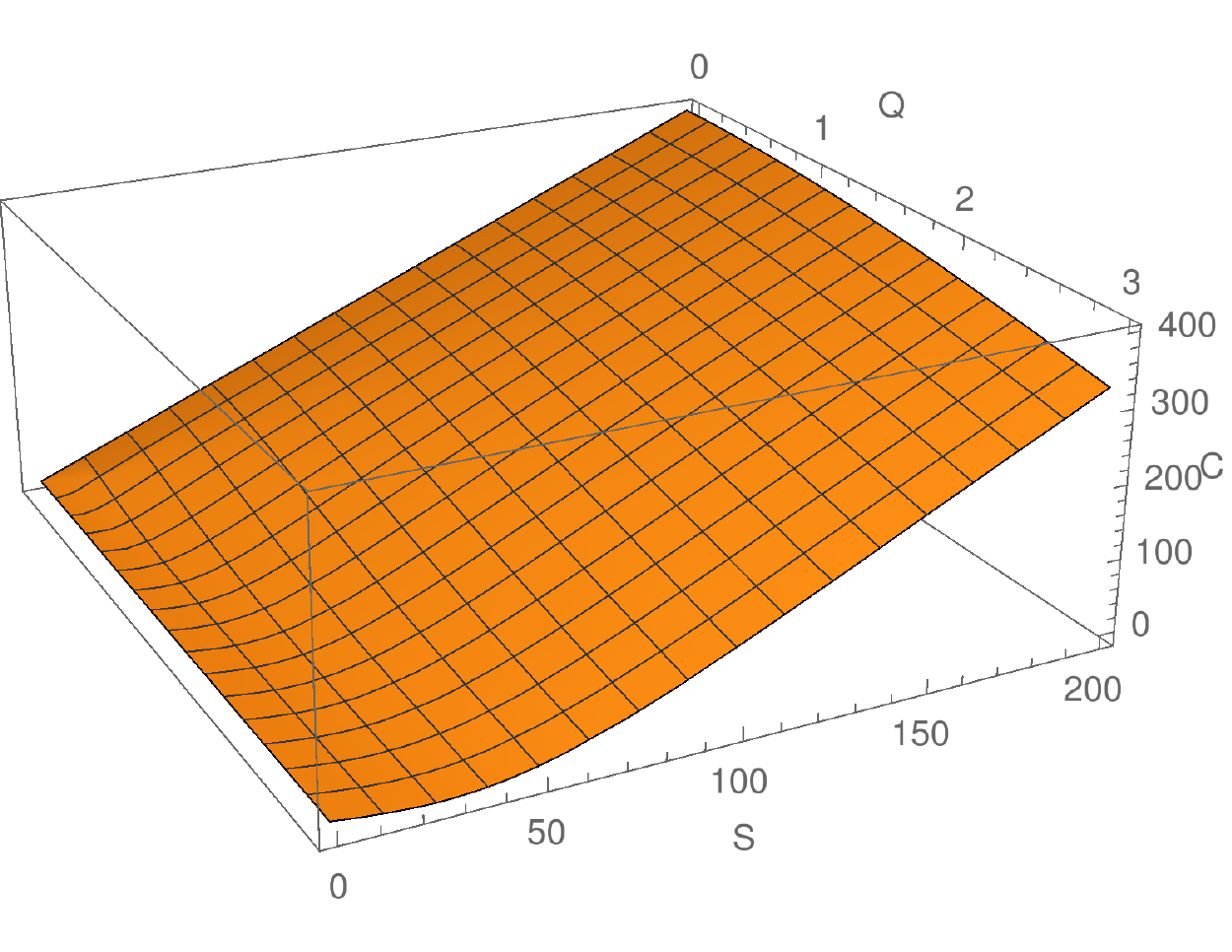}
	      \hspace{1cm}
              \includegraphics[scale=0.4]{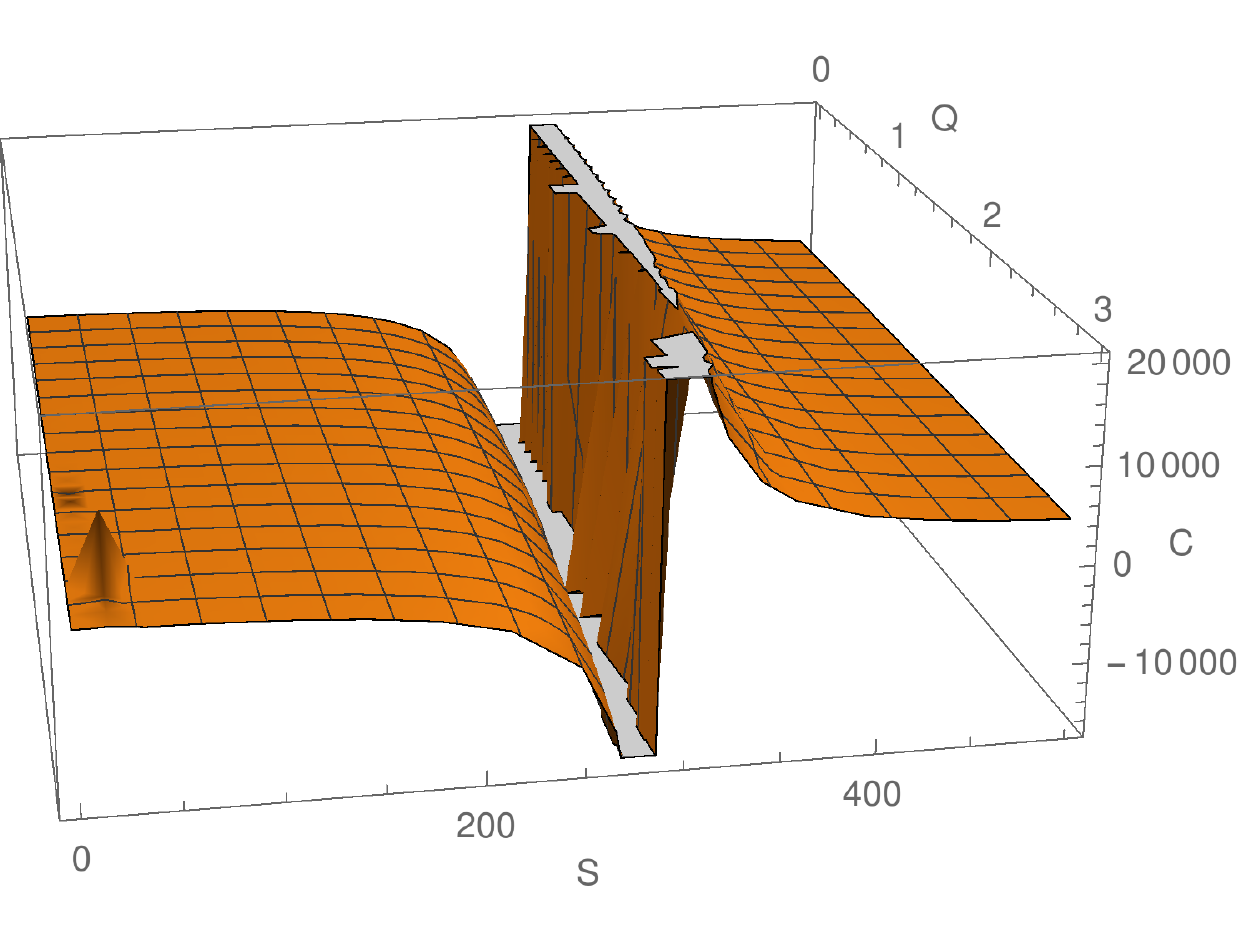}
       	      \hspace{1cm}
              \includegraphics[scale=0.4]{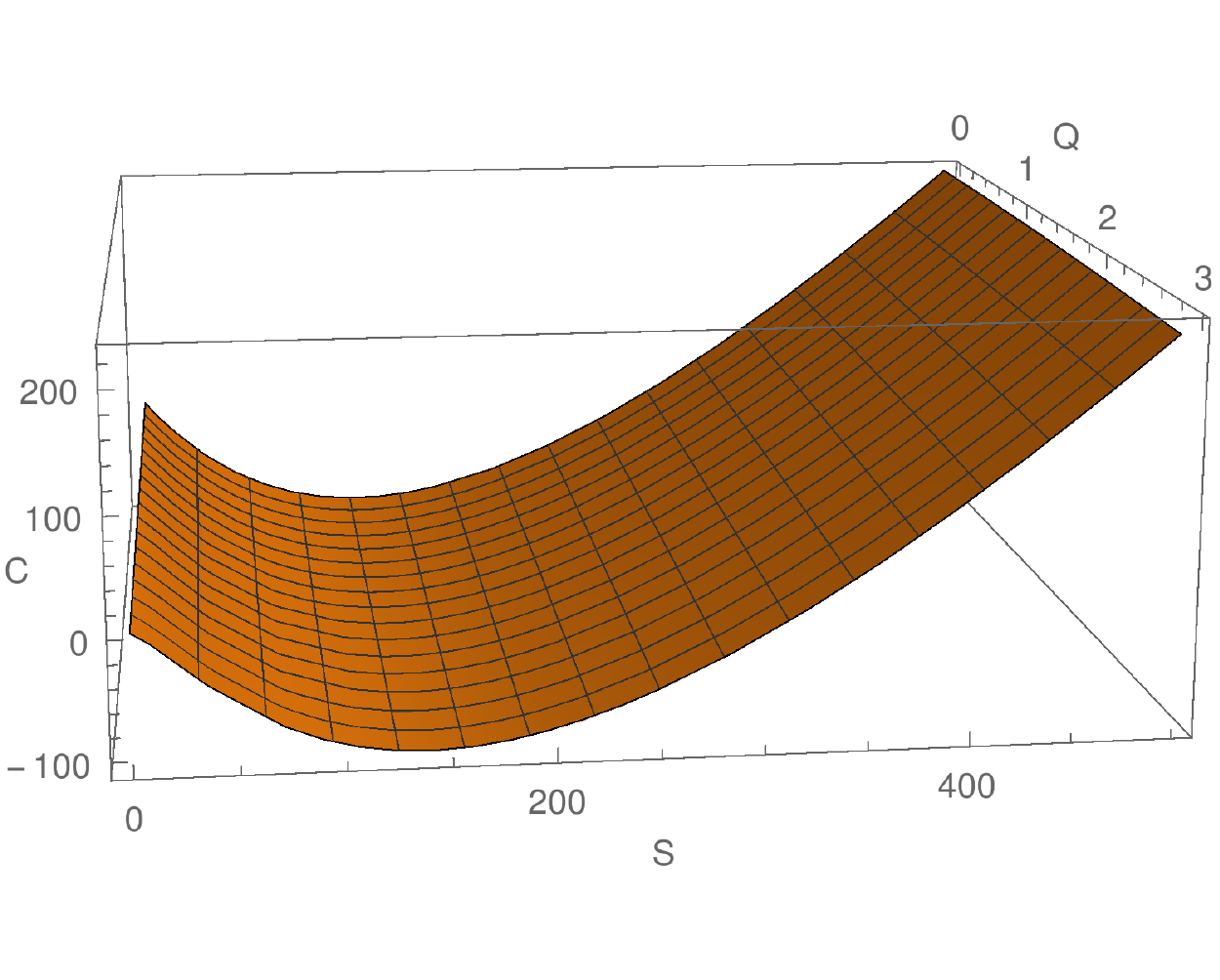}
              \caption{3D Variation of heat capacity against entropy and charge for anti de-Sitter black holes for flat, spherical and hyperbolic topology of space time in massive gravity}
  \label{spec_ads_fig}  
 \end{figure}

In this section we will consider the case in which the curvature parameter is taken as 
negative ($\alpha<0$). For this case the the black hole solution (\ref{MRNds_transformed}) reduces to 
charged anti de Sitter solution in massive gravity (RNAdS). Here also variation of temperature and 
specific heat against entropy for different space time cases are depicted in figure (\ref{temp_ads_fig}) and (\ref{spec_ads_fig}).

Temperature variation in figures (\ref{temp_ads_fig}), implies that for each space time cases, the black hole changes from unphysical region to 
a region which has physical significance. But for the spherically symmetric space time case, the temperature goes to a maximum 
positive value and falls down to lower values. Hence a temperature window-like behaviour is shown by these black holes too.
From the figures (\ref{spec_ads_fig}) we can infer that the for the flat case, the specific heat goes from negative values to positive values and 
hence the black hole system changes from a thermodynamically unstable phase to a stable phase without any 
phase transitions. This behaviour exactly resembles that of RNdS black holes. This resemblance exists
in the hyperbolic space time case too. For the spherically symmetric case, the black hole initially lies in a 
thermodynamically unstable region and transit to a stable phase. Later it undergoes a phase transition in which the stable
black hole becomes an unstable one. 

\subsection{Charged black holes in Einstein's general relativity}
\begin{figure}
              \includegraphics[scale=0.4]{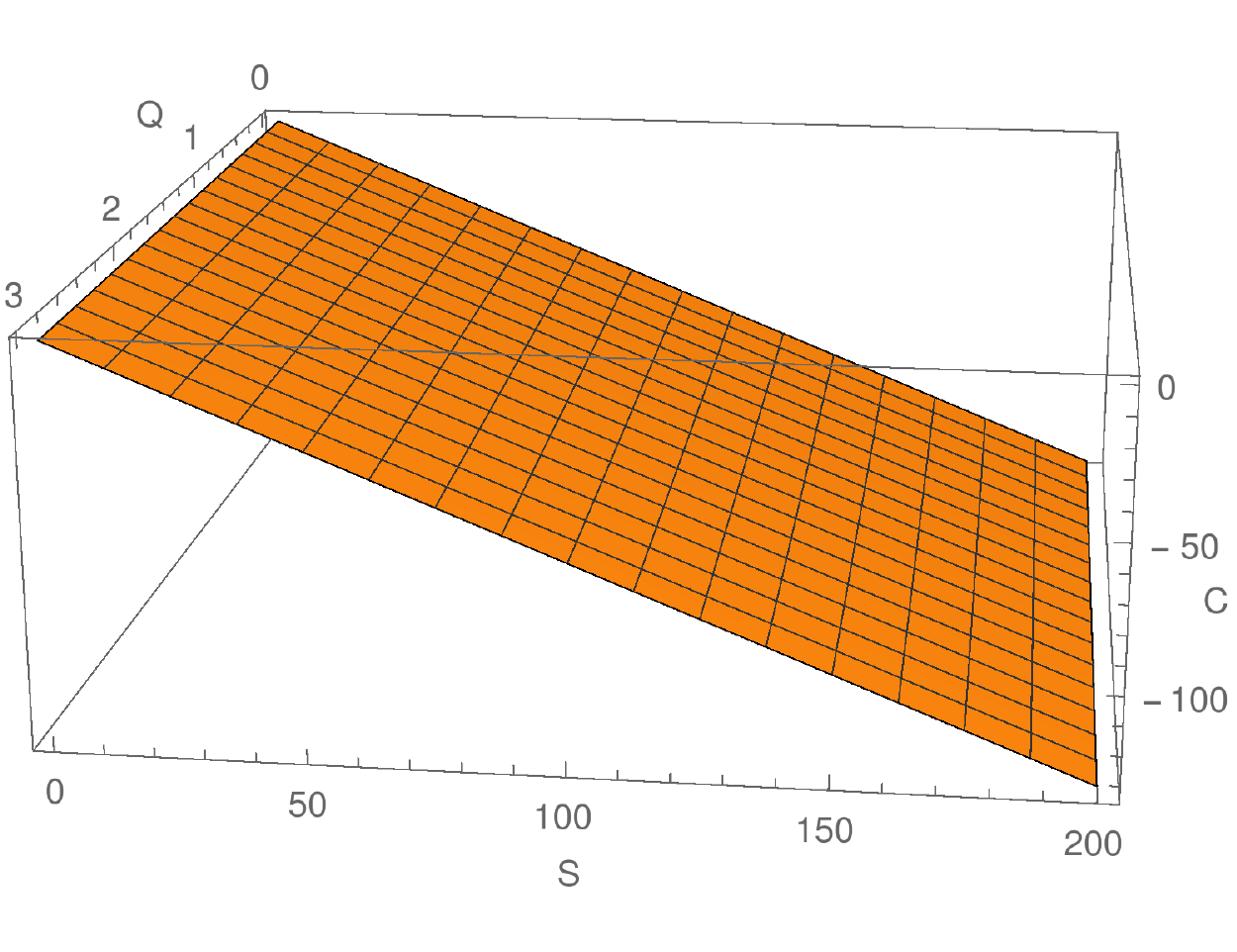}
	      \hspace{1cm}
              \includegraphics[scale=0.4]{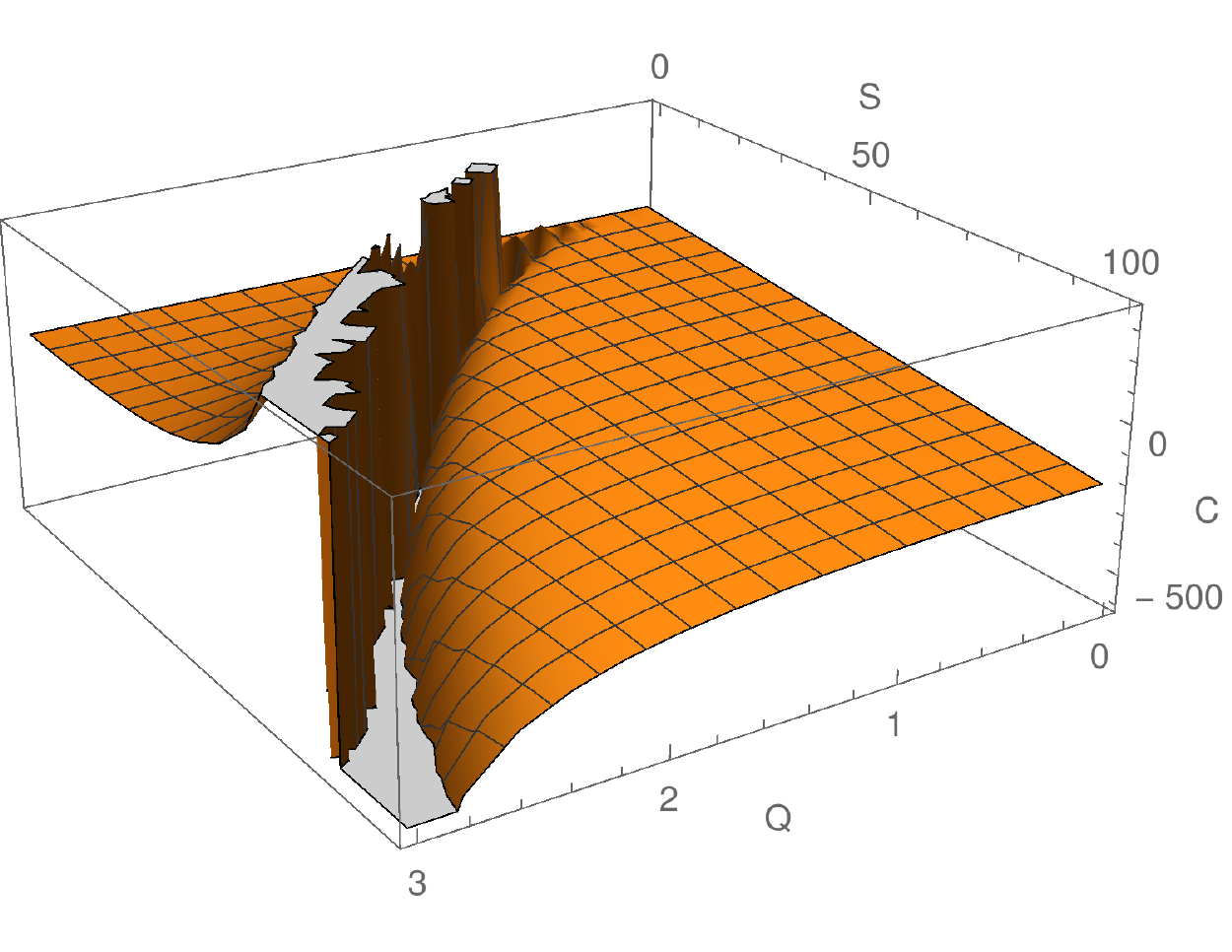}
       	      \hspace{1cm}
              \includegraphics[scale=0.4]{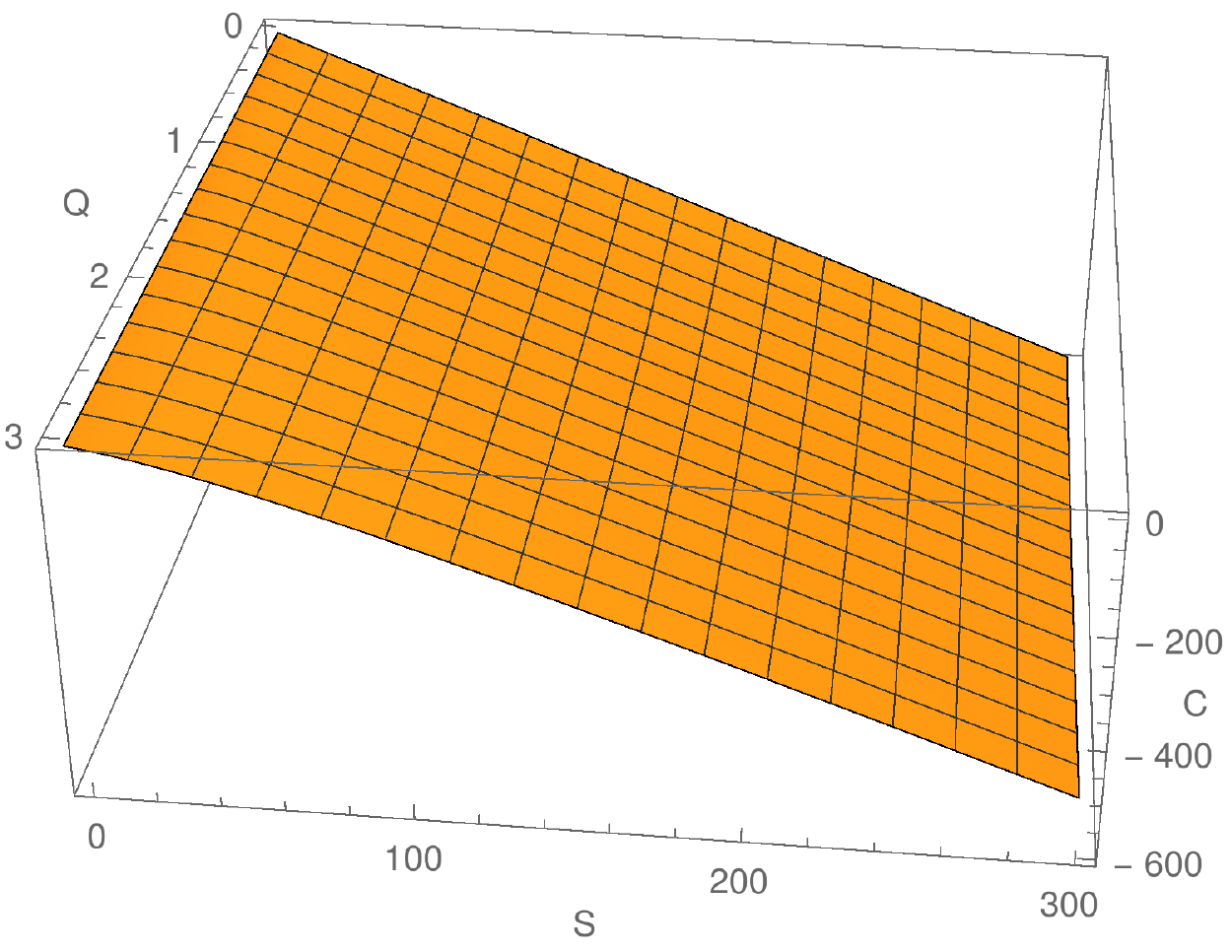}
              \caption{3D Variation of heat capacity against entropy and charge for black holes in Einstein's general relativity for flat, spherical and hyperbolic topology of space time}
   \label{spec_gr_fig} 
 \end{figure}

Now let us consider the situation that the graviton has no mass ($m=0$), then the black hole system given by
(\ref{MRNds_transformed}) will reduce to a charged black hole solution in Einstein's general relativity,
i.e., Reissner-Nordstr\"{o}m 
black hole solution.
For $m=0$ case, one can write down the metric $f(r)$ from (\ref{MRNds_transformed}) as,
\begin{equation}
 f(r)= k - \frac{r_g}{r} + \frac{\tilde{Q}^2}{\tilde{\alpha}^2 r^2}.
 \label{RNmetric}
\end{equation}
 \begin{figure}
              \includegraphics[scale=0.4]{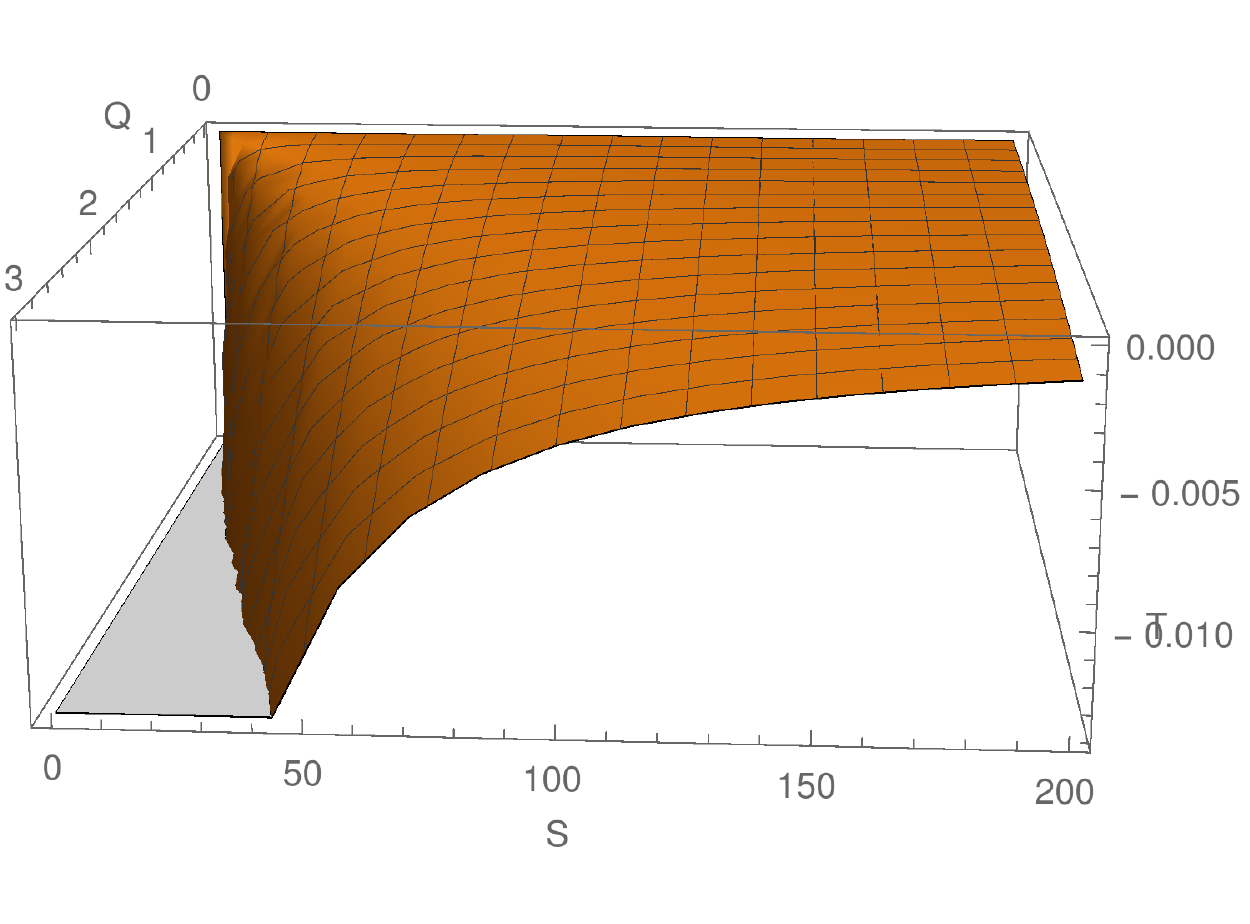}
	      \hspace{1cm}
              \includegraphics[scale=0.4]{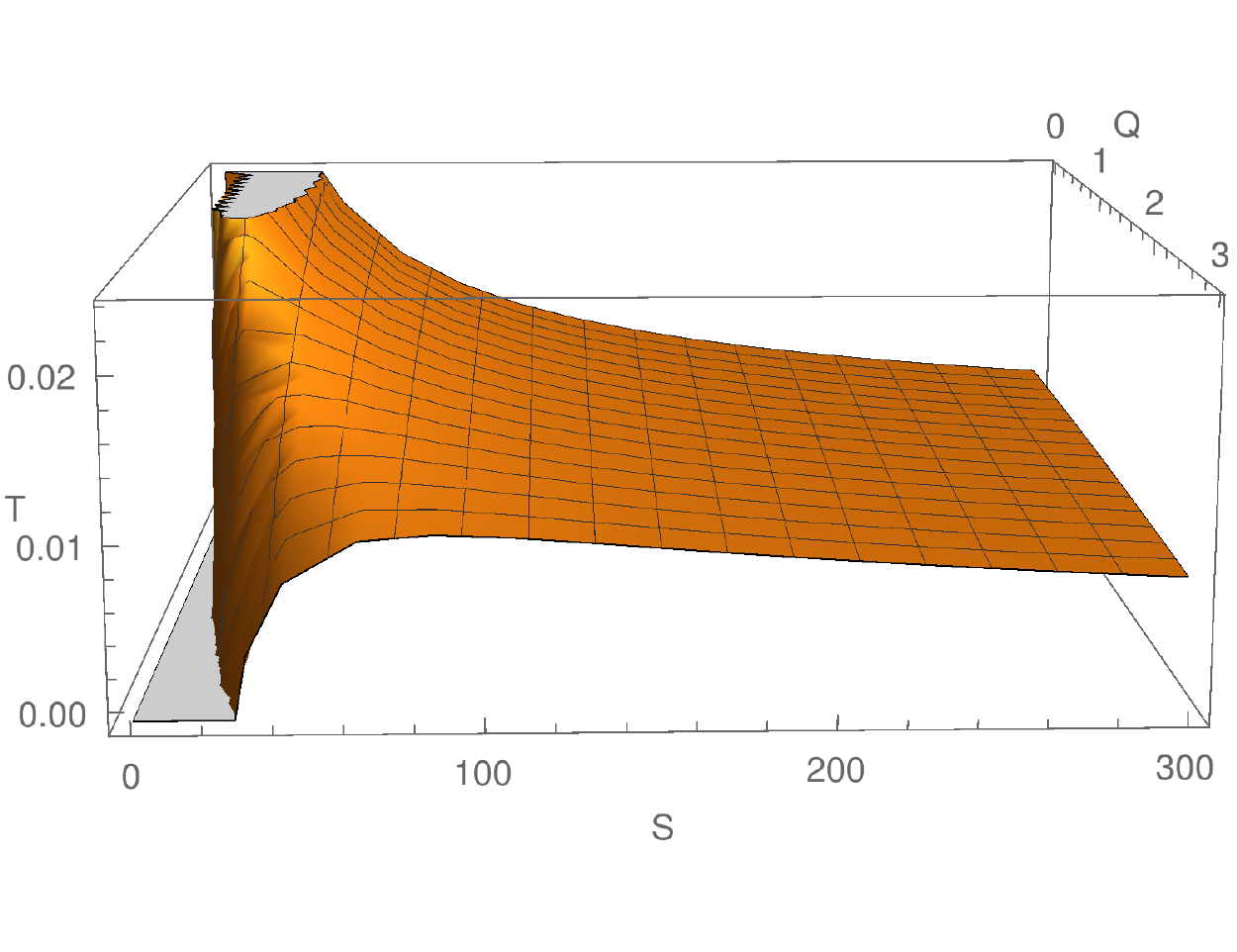}
       	      \hspace{1cm}
              \includegraphics[scale=0.4]{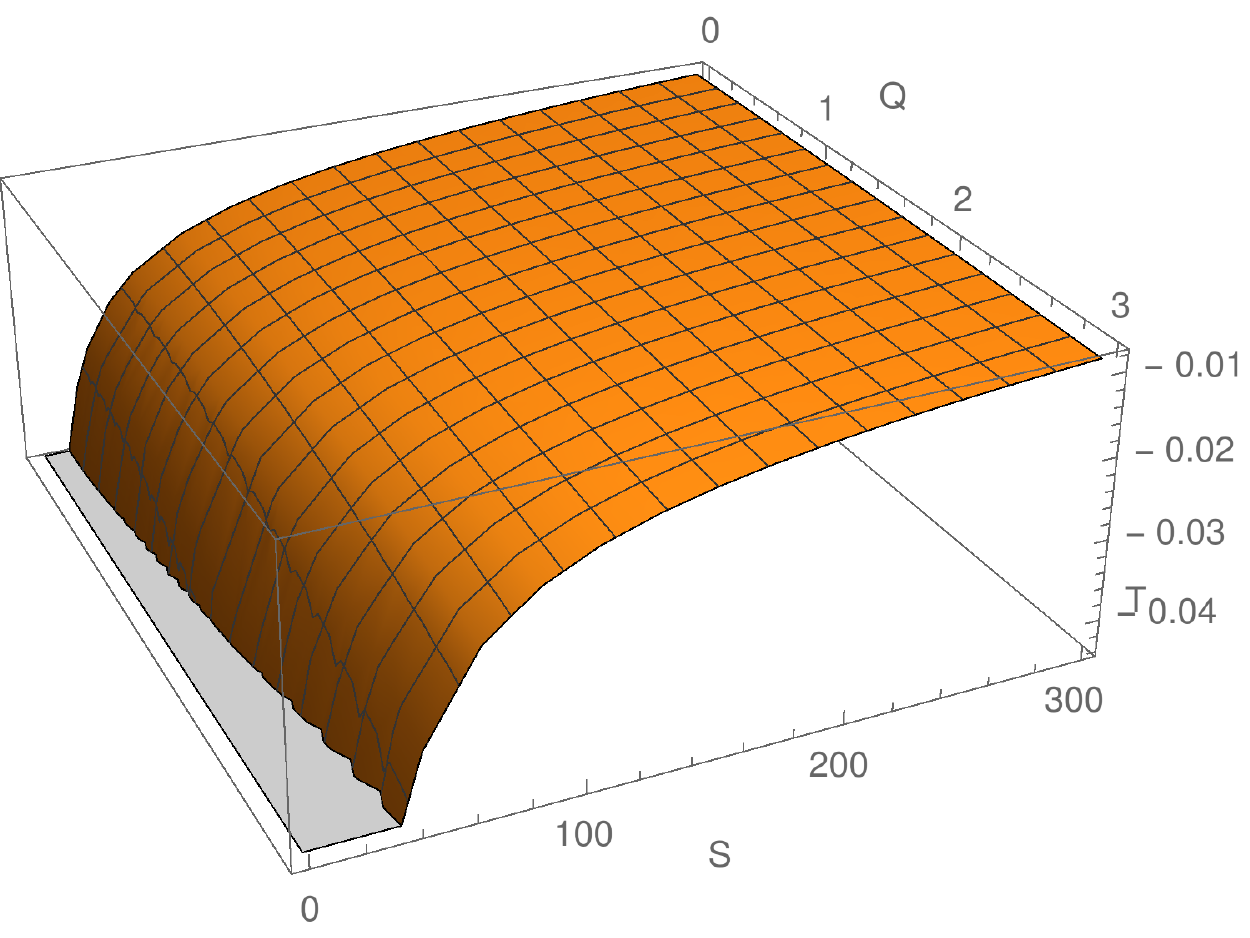}
              \caption{3D Variation of temperature against entropy and charge for black holes in Einstein's general relativity for flat, spherical and hyperbolic topology of space time}
 \label{temp_gr_fig} 
 \end{figure}
It is evident from the above equation that, it exactly matches with the Reissner-Nordstr\"{o}m black hole solution
in Einstein's general relativity. 
Solving the above equation (\ref{RNmetric}), and using the relation $Q \equiv \tilde{Q}/ \tilde{\alpha}$, 
one can easily write the mass of the black hole as,
\begin{equation}
M= \frac{\pi Q^2 +k S} { 2 \sqrt{\pi} \sqrt{S} }.
\end{equation}
From the usual thermodynamic relations $T= \frac{\partial M}{\partial S}$ and $C=T \frac{\partial S}{\partial T}$ one can 
write temperature and heat capacity respectively as,
\begin{equation}
 T=\frac{-\pi Q^2 +k S} { 4 \sqrt{\pi} S^{3/2} },
\end{equation}
and
\begin{equation}
 C= \frac{2S(-\pi Q^2 +k S)} { 3\pi Q^2 -k S }.
\end{equation}
The variation of both temperature and heat capacity are plotted. From the figures (\ref{temp_gr_fig}) and (\ref{spec_gr_fig}), we can see that
for flat case ($k=0$) as well as for the hyperbolic ($k=-1$) case temperature always lies in the negative value region
and hence in physically insignificant region. Now for the spherically symmetric space time case ($k=1$), black hole initially
lies in a negative temperature region and as the black hole horizon radius increases it goes to a
maximum temperature value. After the maximum value of temperature is attained it lies in the positive temperature 
region itself. We can see that
when mass of the graviton
becomes zero, it exactly reproduces the results of RN black holes. So the limiting case of RNdS or RNAdS black hole in massive gravity
coincides with RN solution in Einstein's theory and their thermodynamic behaviour gives the proof for the same.
\section{Geometrothermodynamics of charged black holes in massive gravity}

To investigate the mathematical structure of thermodynamic systems, it is very convenient to use contact geometry. 
In order to incorporate the idea of differential geometry, one has to first define a $2n+1$ thermodynamic phase space $(\mathcal{T})$. 
This phase space is coordinated by the set $Z^a= \{ \Phi,E^a , I^a \}$, where $\Phi$ is an arbitrary thermodynamic
potential, the coordinates $E^a$ correspond to the extensive variables and $I^a$ are the corresponding dual
intensive variables, where $a=1,...,n.$ We can introduce the contact 1-form or fundamental Gibbs 1-form as \cite{gibbs},
\begin{equation}
 \Theta = d\Phi- \delta_{ab} I^a dE^b, ~~ \delta_{ab}=\mbox{diag}(1,1,,..,1),
\end{equation}
with summation over the repeated indices. Now if $\mathcal{T}$ is differentiable and $\Theta$ satisfies the condition
$\Theta \wedge (\mbox{d} \Theta)^{n} \neq 0$, then $(\mathcal{T},\Theta)$ can be called as a contact manifold. 
Now consider an $n$ dimensional manifold $\mathcal{E}$, which is a sub manifold of  $\mathcal{T}$, i.e., $\mathcal{E} \subset \mathcal{T}$, 
which can be defined using the extensive thermodynamic variables $E^a$.
This equilibrium manifold $\mathcal{E}$ can be realized by considering a smooth harmonic mapping 
 $\varphi : \mathcal{E} \rightarrow \mathcal{T}$,
  \begin{equation}
  \varphi : \mathcal{E^a} \rightarrow \{ Z^a (E^a)\}=\{ \Phi(E^a),E^a , I^a(E^a)  \},
 \end{equation}
and the condition $\mathcal{E} \subset \mathcal{T}$, can be realized using the condition,
\begin{equation}
\varphi^* (\Theta) = \varphi^* (\mbox{d}\Phi- \delta_{ab} I^a dE^b) =0,
\label{pullback}
\end{equation}
where $\varphi^*$ is the pullback. From the relation (\ref{pullback}) one can easily deduce the 
condition for thermodynamic equilibrium as,
\begin{equation}
 \frac{\partial \Phi}{\partial E^a} = \delta_{ab} I^b.
\end{equation}
Considering the equilibrium manifold $\mathcal{E}$ and using (\ref{pullback}), the first law of thermodynamics can
be written as,
\begin{equation}
\varphi^*(\Theta)=\varphi^*(d\Phi -\delta_{ab}\, I^a\, dE^b) = 0\ .
\end{equation}
The harmonic map $\varphi$ demands the existence of the function $\Phi=\Phi(E^a)$, which is commonly known as the fundamental 
equation in classical thermodynamics from which one can deduce all the equation of states corresponding to that system. From this
fundamental equation, one can write the second law of thermodynamics as,
\begin{equation}
 \pm \frac{\partial^2 \Phi}{\partial E^a  \partial E^b} \geq 0 ,
\end{equation}
also known as the convexity condition. In the above equation, the sign depends on the choice of 
the thermodynamic potential. For example, if one chooses 
$\Phi$ as entropy, then the sign becomes positive and it becomes negative when the potential is chosen 
to be the internal energy. 
Now let us consider a Riemannian metric
$G$ on $\mathcal{T}$, which must be invariant with respect to Legendre transformations. Then the Riemannian 
contact manifold can be defined
as the set $(\mathcal{T},\Theta,G)$ and the equilibrium manifold can be written as a sub
manifold of $\mathcal{T}$, i.e., $\mathcal{E} \subset \mathcal{T}$.
This sub manifold satisfies the above discussed pull back condition \cite{jackle}. 
The non-degenerate metric $G$ and the thermodynamic metric $g$ can be written as,
\begin{equation}
 G=(d\Phi - \delta_{ab} I^a d E^b)^2 +(\delta_{ab} E^a I^b)(\eta_{cd} d E^c d I^d),
\end{equation}
and,
\begin{equation}
  g^Q=\varphi^*(G)=\left(E^{c}\frac{\partial{\Phi}}{\partial{E^{c}}}\right)
\left(\eta_{ab}\delta^{bc}\frac{\partial^{2}\Phi}{\partial {E^{c}}\partial{E^{d}}} dE^a dE^d \right),
\label{quevedo metric}
\end{equation}
with $\eta_{ab}$=diag(-1,1,1,..,1) and this metric is Legendre invariant because of the invariance of the Gibbs 1-form.
Now by calculating the curvature scalar of the GTD metric(\ref{quevedo metric}), one can use GTD as a method to investigate
the phase transition structure of the black hole system. 

\begin{figure}
              \includegraphics[scale=0.4]{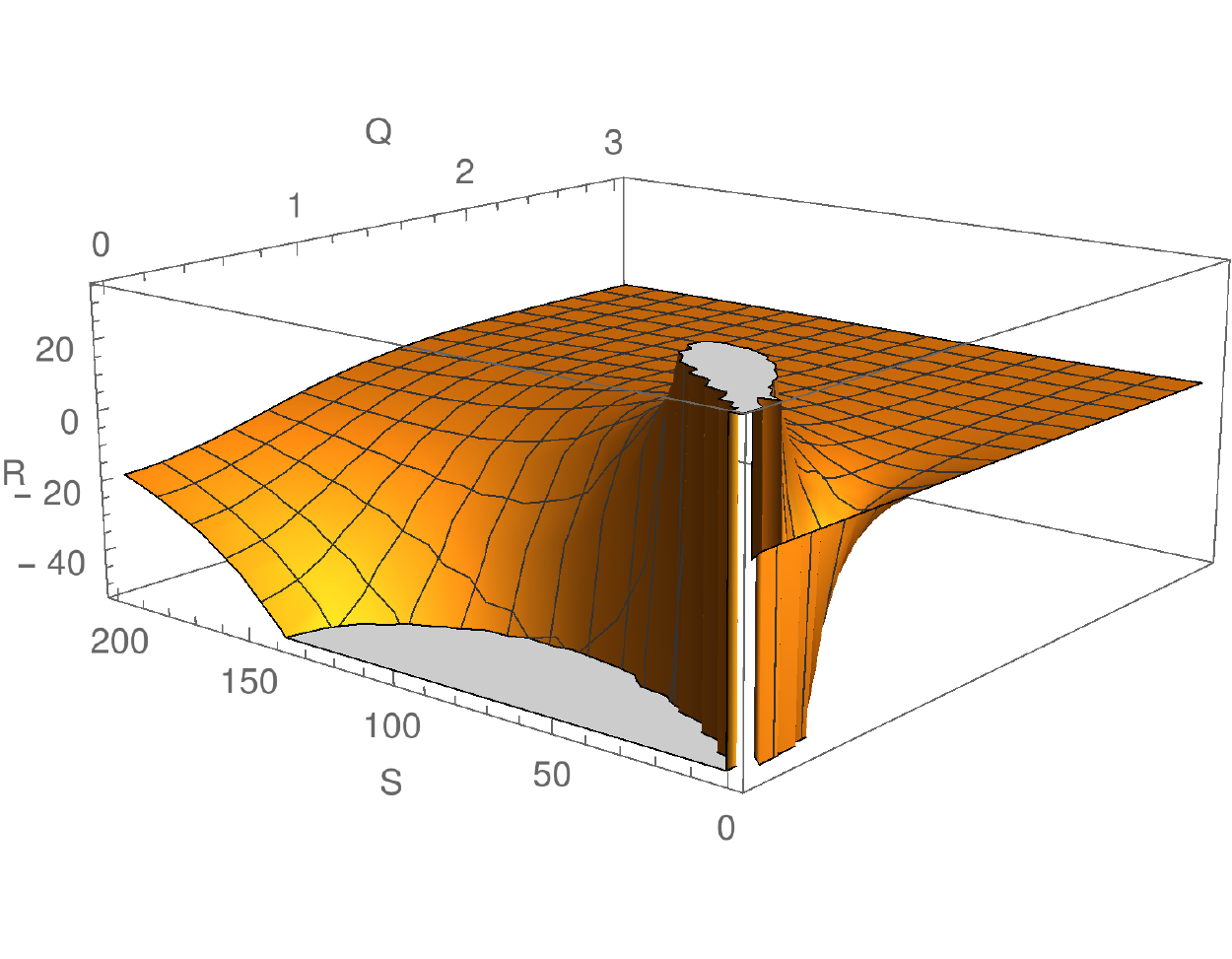}
	      \hspace{1cm}
              \includegraphics[scale=0.4]{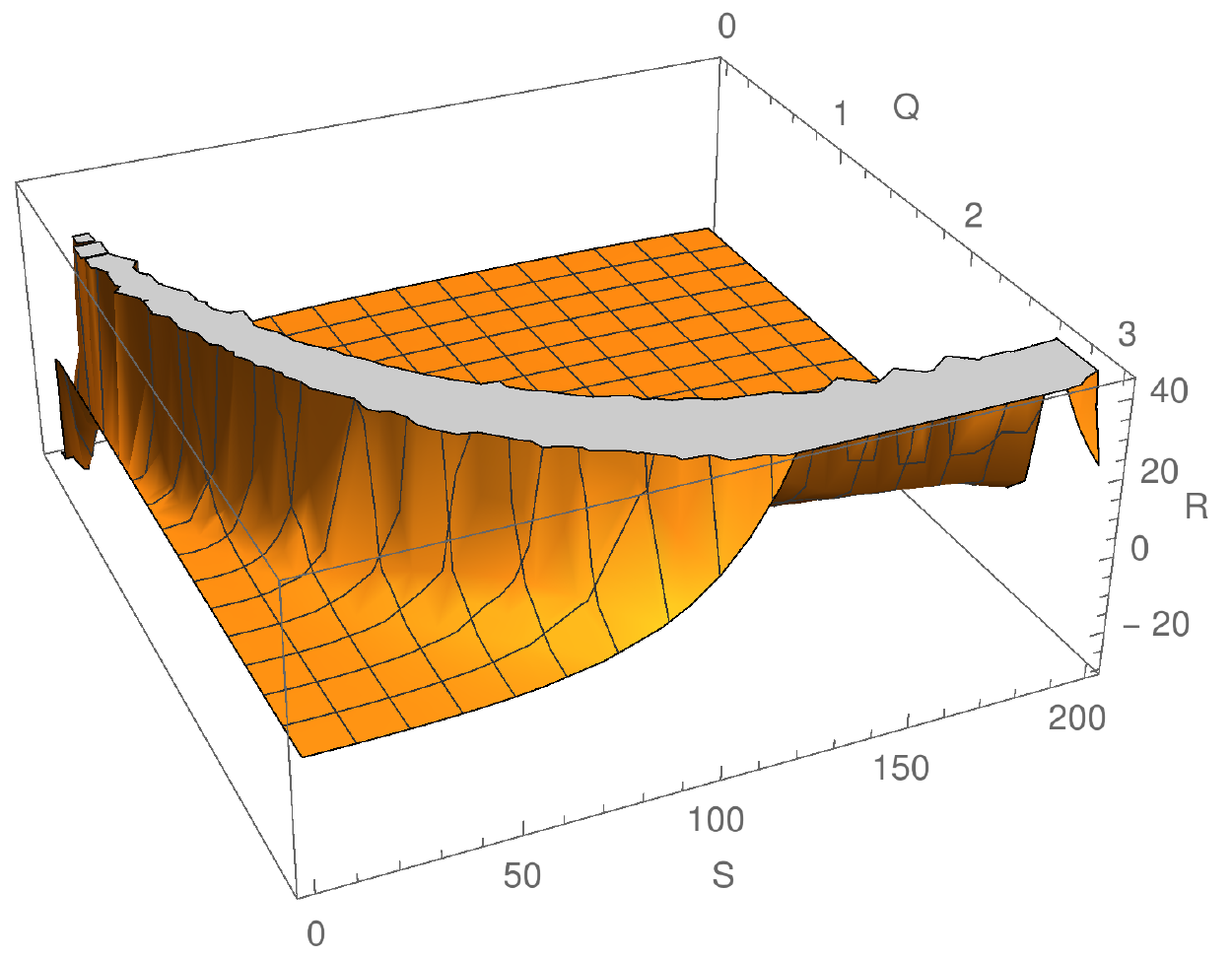}
       	      \hspace{1cm}
              \includegraphics[scale=0.4]{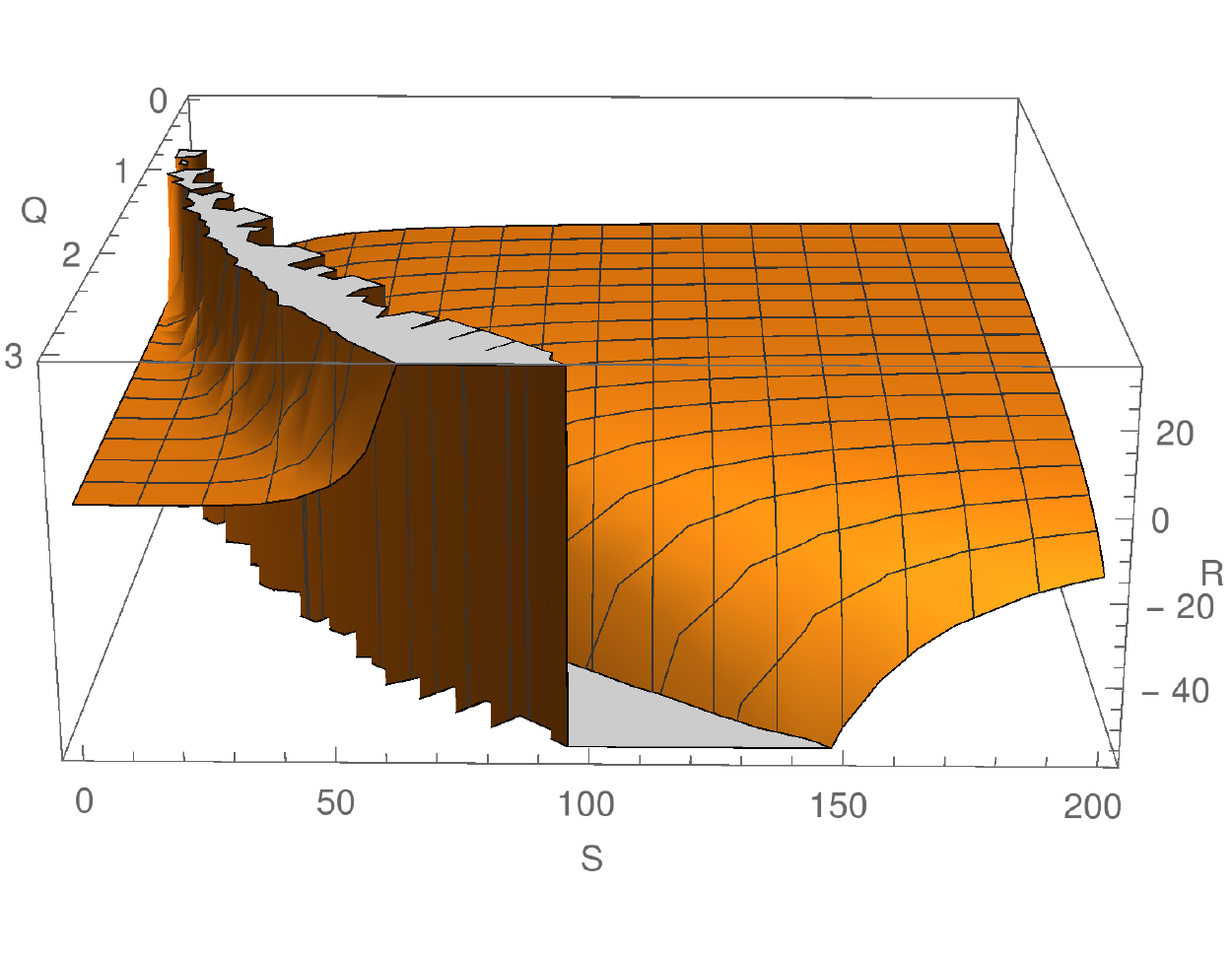}
              \caption{3D Variation of Ricci scalar against entropy and charge anti de-Sitter black holes for for flat, spherical and hyperbolic topology of space time in massive gravity}
 \label{ricci_ads_fig}  
 \end{figure}

\begin{figure}
              \includegraphics[scale=0.4]{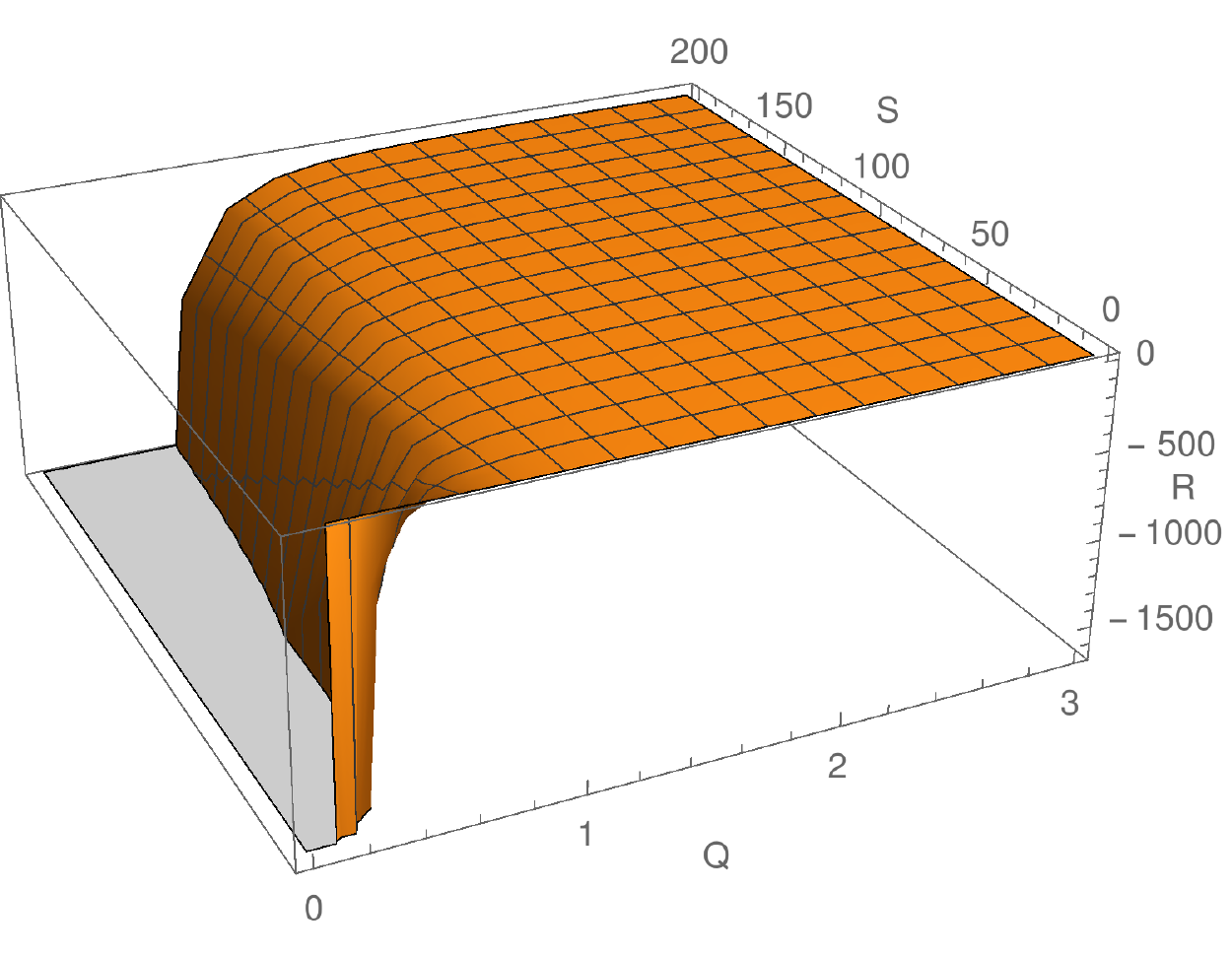}
	      \hspace{1cm}
              \includegraphics[scale=0.4]{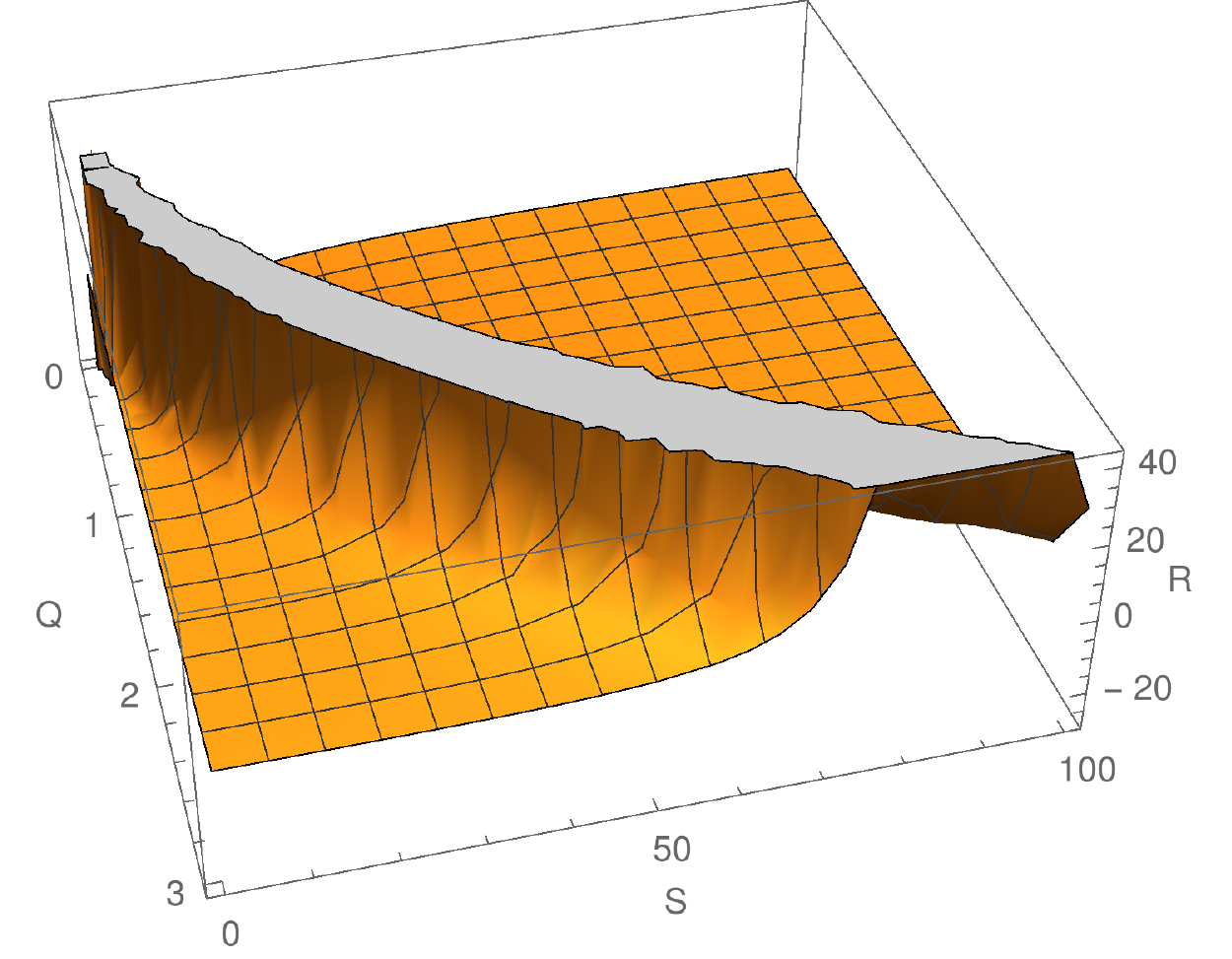}
       	      \hspace{1cm}
              \includegraphics[scale=0.4]{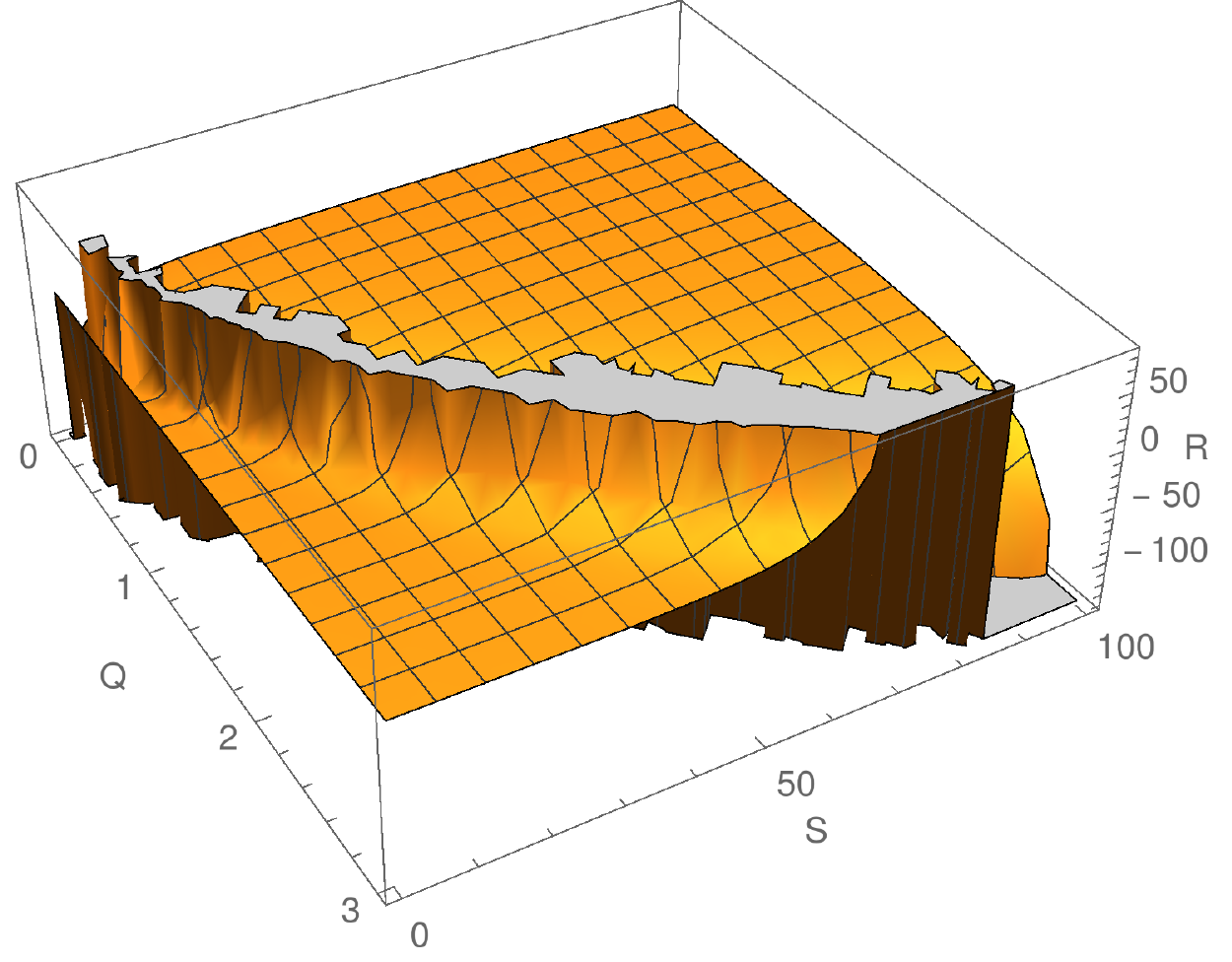}
              \caption{3D Variation of Ricci scalar against entropy and charge de-Sitter black holes for for flat, spherical and hyperbolic topology of space time in Einstein's gravity}
   \label{ricci_grds_fig}  
 \end{figure}

\begin{figure*}
              \includegraphics[scale=0.4]{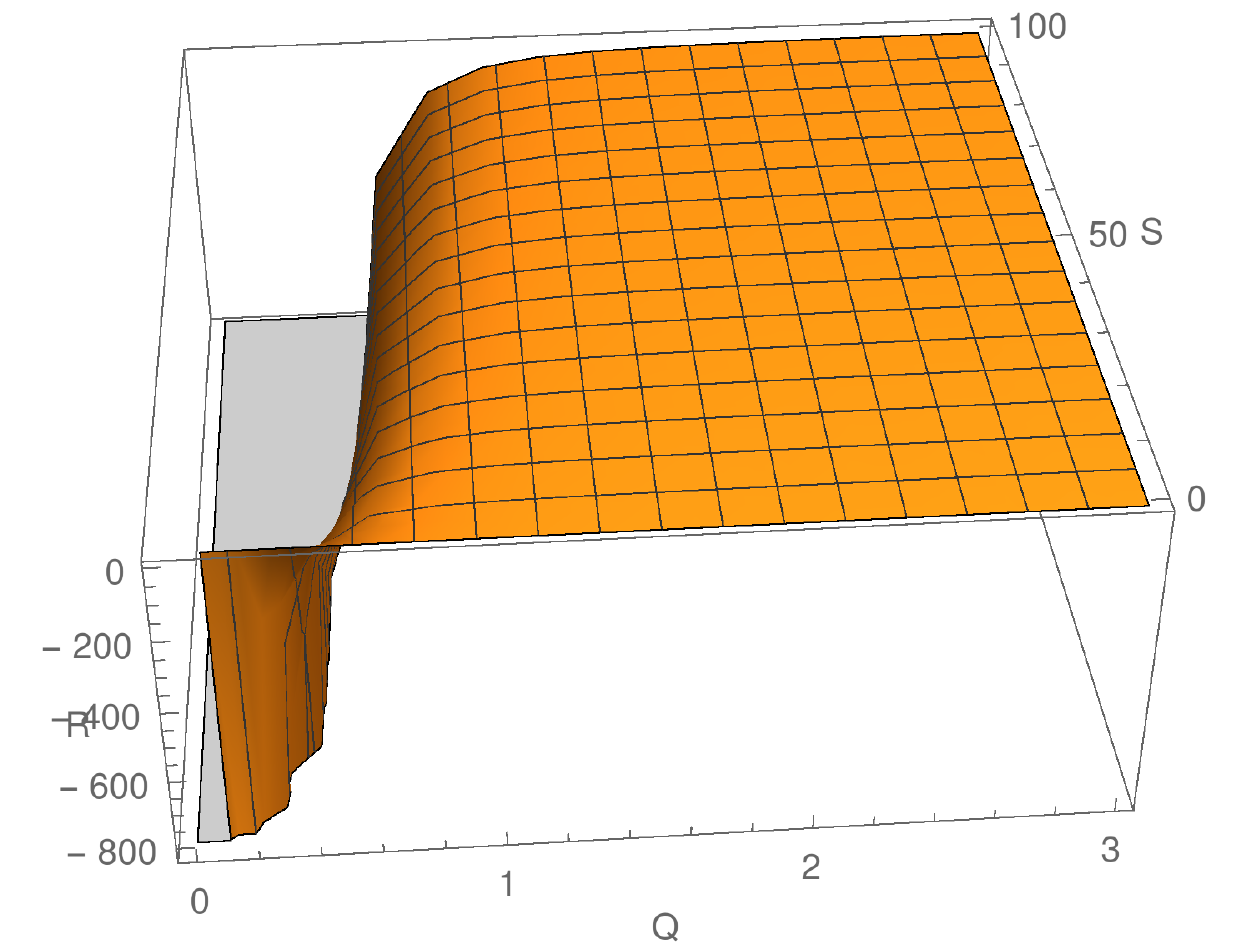}
	      \hspace{1cm}
              \includegraphics[scale=0.4]{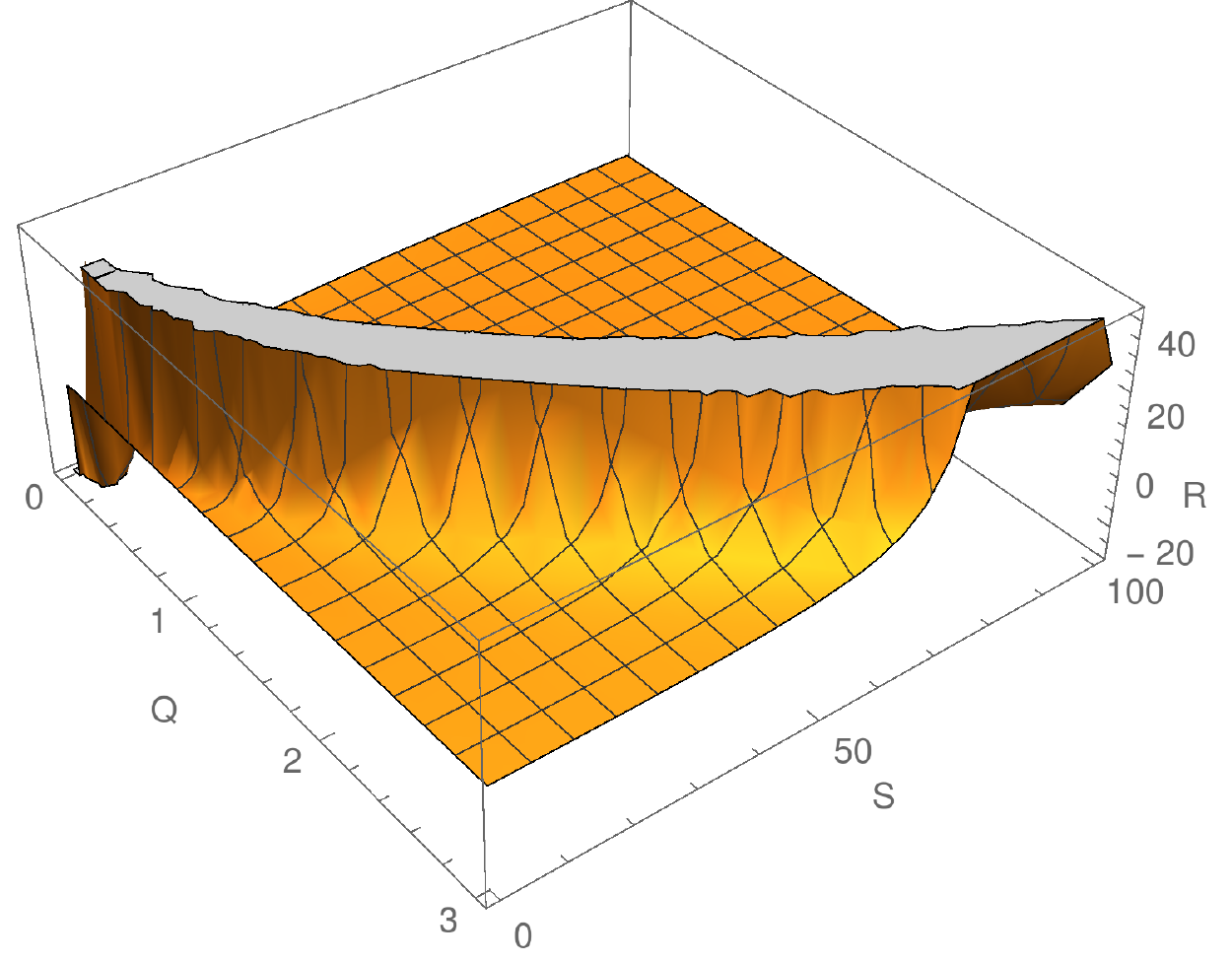}
       	      \hspace{1cm}
              \includegraphics[scale=0.4]{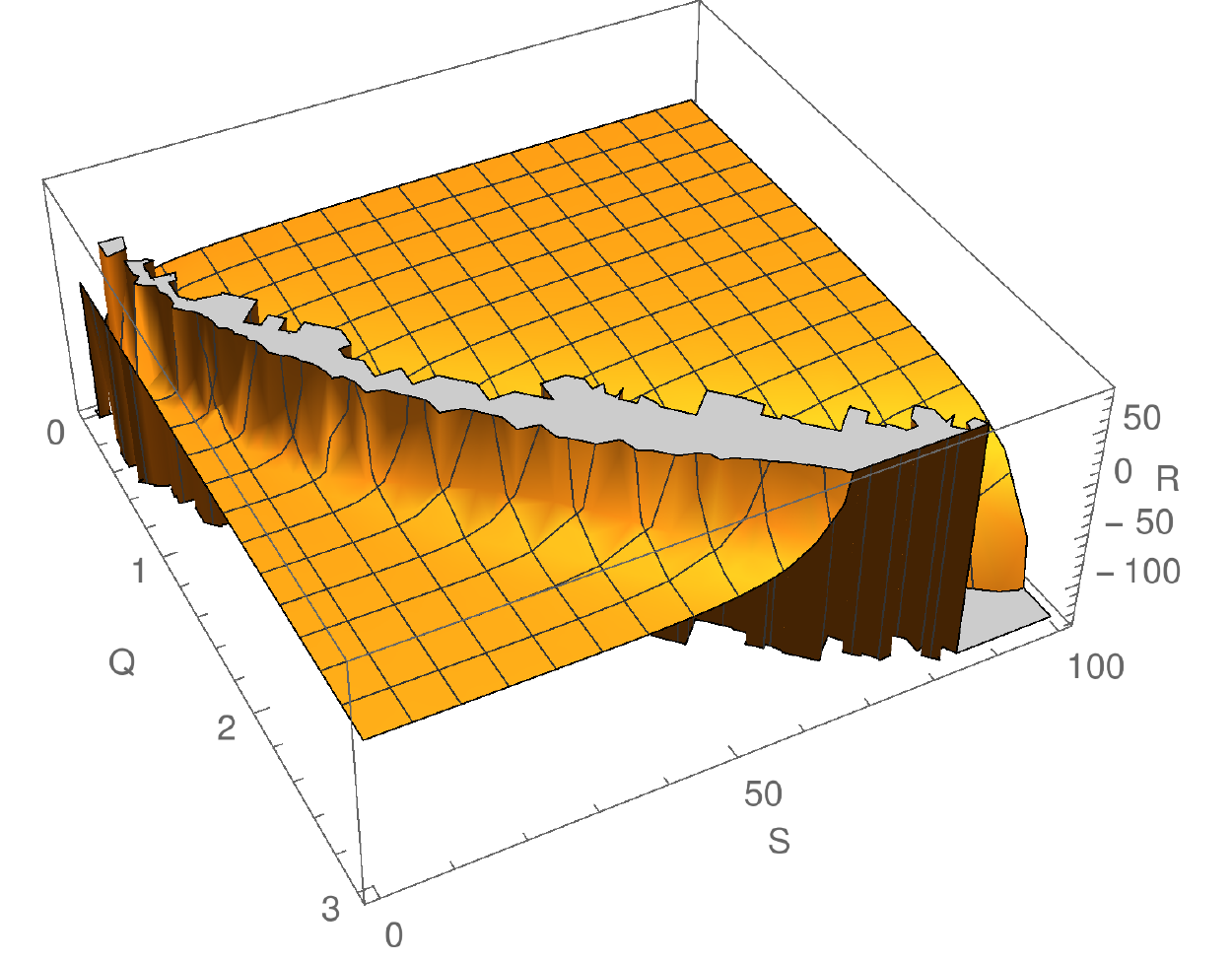}
              \caption{3D Variation of Ricci scalar against entropy and charge anti de-Sitter black holes for for flat, spherical and hyperbolic topology of space time in Einstein's gravity}
   \label{ricci_grads_fig}  
 \end{figure*}

In this section we investigate the phase transition of the charged black hole solution in massive gravity by using the concept of 
Geometrothermodynamics (GTD). Now we will introduce the ideas of GTD in to the charged black hole system in
massive gravity. For this, we will construct a $7$ dimensional thermodynamic phase space $\mathcal{T}$ using the 
extensive variables and their dual intensive variables as coordinates. In the present case, coordinates are given by
$Z^A=\{ M, S, Q, \alpha , T, \phi, a \} $, where $S, Q, \alpha$ are extensive variables and $T, \phi, a$ are their corresponding
dual intensive variables. Now one can write the Gibbs 1-form as,
\begin{equation}
 \Theta= dM-TdS-\phi dQ - a d\alpha.
\end{equation}
Now using the idea of equilibrium manifold, one can write the GTD metric as,
\[
         g=(SM_{S}+Q M_{Q}+ \alpha M_{\alpha})
            \left[ {\begin{array}{ccc}
             -M_{SS} & 0 & 0  \\
             0 &  M_{QQ} & M_{Q \alpha} \\
             0 & M_{\alpha Q} & M_{\alpha \alpha} \\
             \end{array} } \right].
        \]
Then, the Legendre invariant scalar curvature corresponding to the above metric is given by,
\begin{equation}
 R^{GTD}= \frac{f(S,Q,\alpha)}{\left(\pi  \alpha  \left(k S-3 \pi  Q^2\right)
	  +2 m^2 S^2\right)^2 \left(2 m^2 S^2-3 \pi  \alpha  \left(k S+3 \pi  Q^2\right)\right)^3}
\end{equation}
where $f(S,Q,\alpha)$ is a complicated expression of less physical interest. Now we will investigate the thermodynamic
behaviour of the black hole system using the scalar curvature. According to the theory of Geometrothermodynamics, 
the points of zero scalar curvature as well as the infinite discontinuities will exactly matches with the 
singular behaviours of thermodynamic potentials which corresponds to the black hole system. 
Now let us evaluate different charged black hole system in massive gravity, as they vary with respect to 
mass of the graviton, topology of the solutions and the sign of the curvature parameter. 
 \begin{figure}
              \includegraphics[scale=0.4]{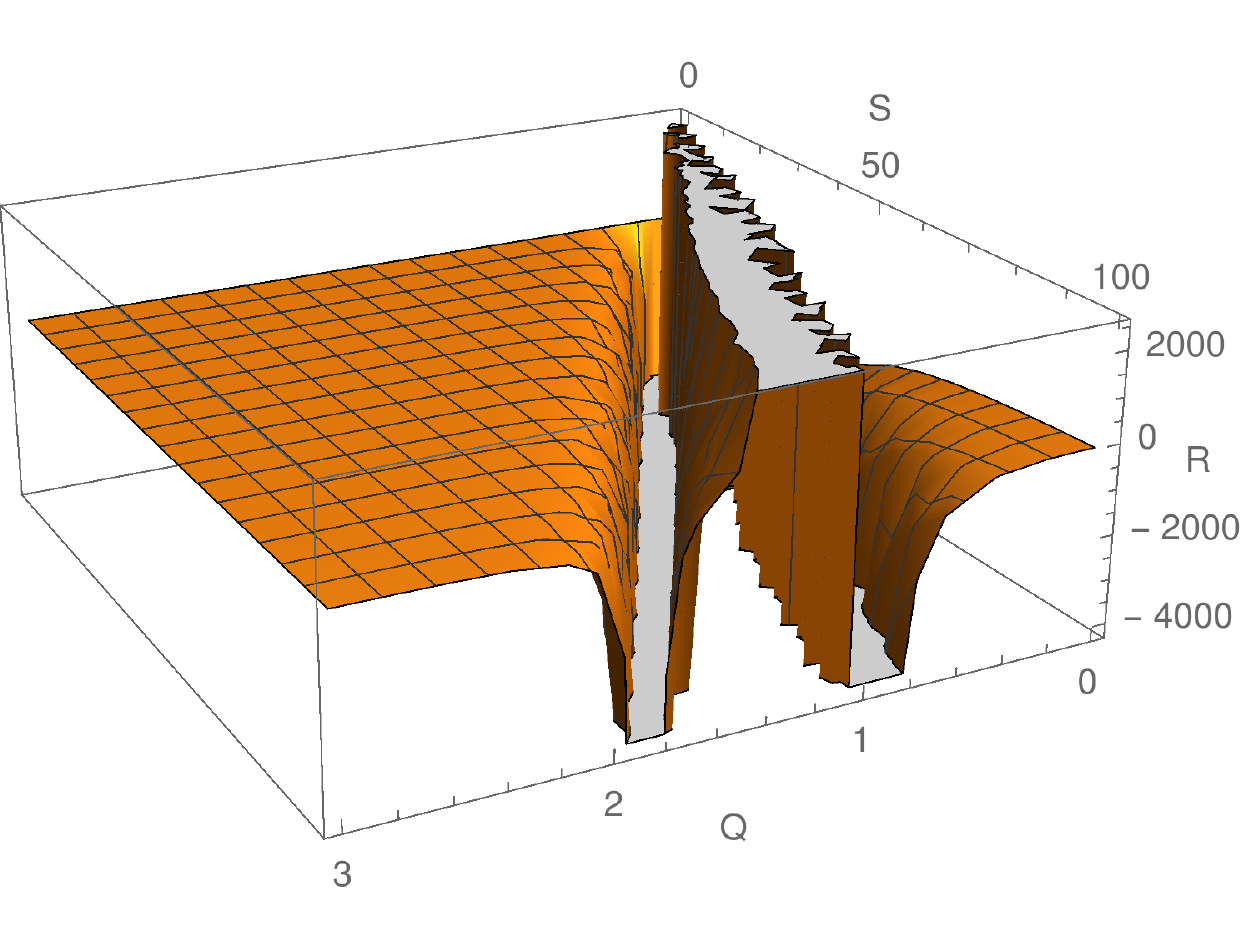}
	      \hspace{1cm}
              \includegraphics[scale=0.4]{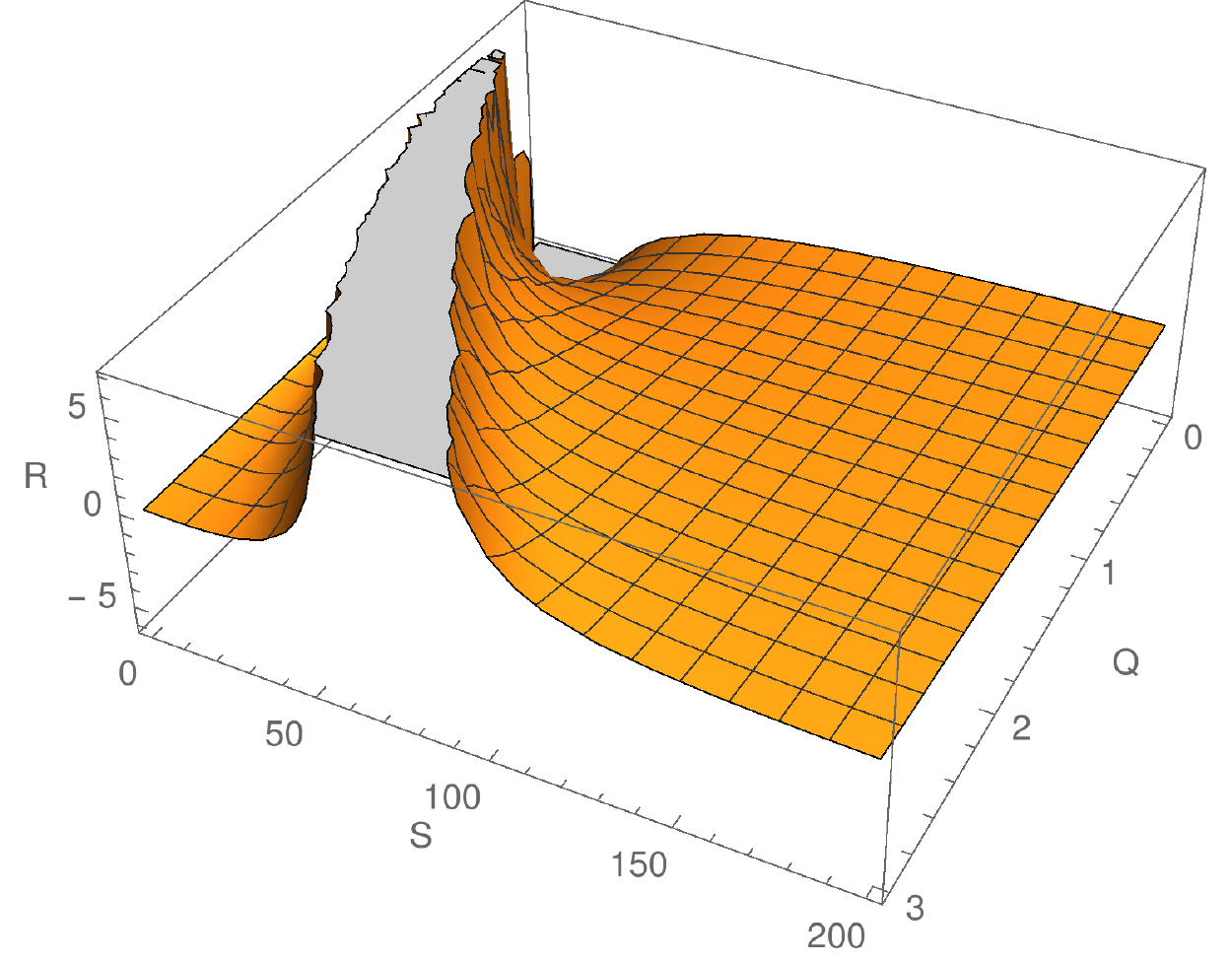}
       	      \hspace{1cm}
              \includegraphics[scale=0.4]{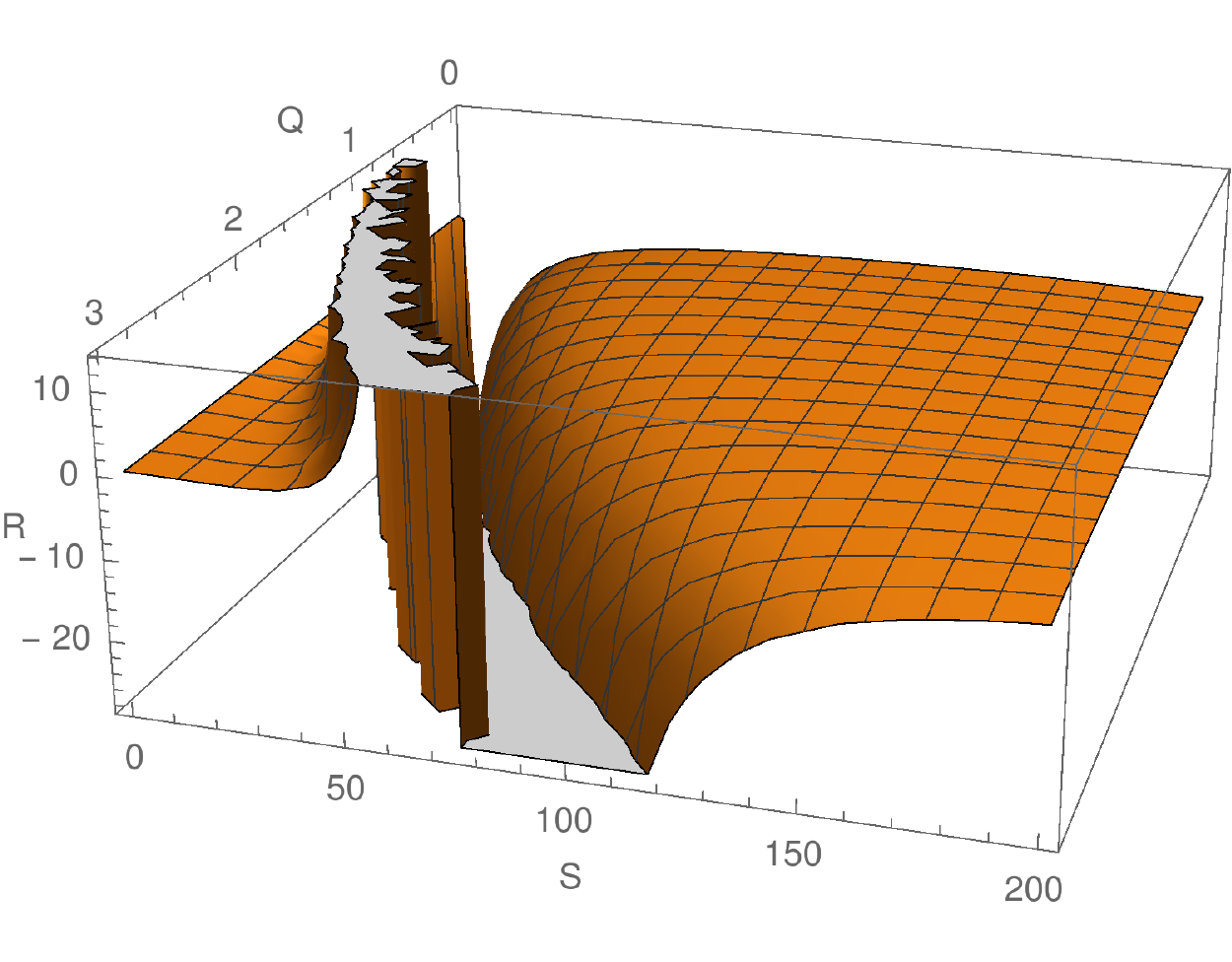}
              \caption{3D Variation of Ricci scalar against entropy and charge de-Sitter black holes for for flat, spherical and hyperbolic topology of space time in massive gravity}
  \label{ricci_ds_fig}  
 \end{figure}
 The variation of Ricci scalar curvature with entropy is depicted in figures (\ref{ricci_ds_fig}) and (\ref{ricci_ads_fig}). 
Let us first consider the case in which the curvature parameter, $\alpha$ is taken as positive, then as we have discussed 
earlier, the black hole system will behave like charged de-Sitter black hole system. From the figures (\ref{ricci_ds_fig}) it is evident that,
all singular points in the thermodynamic parameters like temperature and heat capacity are exactly reproduced by
the Ricci scalar either by vanishing or showing infinite discontinuities at the same points. 
For the second case also Ricci scalar behaves in a similar manner, where the curvature parameter, $\alpha$ is taken as
negative, and hence the the black hole system becomes a charged anti de-Sitter black hole system. The variation of scalar curvature
for this case is plotted in figures (\ref{ricci_ads_fig}). 
Now let us investigate the Geometrothermodynamics and the behaviour of Legendre invariant scalar curvature of the black
hole system, when the mass off the graviton becomes zero. The corresponding variation of Ricci scalar is depicted in 
figure (\ref{ricci_grds_fig}) and (\ref{ricci_grads_fig}). Here too, the singularities of the Ricci scalar matches with those of temperature and heat capacity of the RN 
black hole in Einstein's general relativity. It is interesting to note that the GTD results obtained from the 
charged black hole solution in massive gravity coincides with the study of RN black hole 
previously obtained in \cite{sanchez,rnq}. 
Hence the Geometrothermodynamics exactly reproduces the phase transition structure of the charged black hole solutions
in massive gravity.  
 
\section{Results and Discussion}
In this paper, we have studied the black hole thermodynamics and geometrothermodynamics of the charged de-Sitter and anti de-Sitter
solutions in massive gravity. The main aim of this paper is to discuss the phase transition as well as the singular behaviours
in thermodynamic potentials using GTD method. We used the Quevedo metric or GTD metric to obtain the 
Legendre invariant Ricci scalar curvature. We analysed the thermodynamic behaviour using both analytical and graphical methods. 
The analysis showed that, the singular behaviours in the thermodynamic potentials, including the point where 
heat capacity diverges, exactly reproduced by the Ricci scalar obtained using the GTD metric. Hence we can say that, 
GTD metric exactly reproduces the phase transition structure of 
charged black holes in massive gravity and their corresponding thermodynamic interactions. From this
study it is evident that the charged black holes in massive gravity undergoes a phase transition. One expects that thermodynamics of black holes would be the same as in general relativity, taking into account that massive gravity differs from general relativity by a non-derivative coupling to a fiducial metric. But our studies show that, even though the results agrees with general relativity when massive parameter tends to zero, there are significant changes in the phase transition structure of the system when $m\neq0$. Comparative studies on analytical and graphical representation of the changes of heat capacity and Ricci scalar against entropy and massive parameter of the black hole system reveal the same result.   Also the present
study shows that, like all other charged black hole solutions in Einstein's gravity and in all modified gravities, 
there exists a temperature window, where the black hole temperature lies in a physically significant region with 
positive temperature.

\section{Acknowledgement}
The authors wish to thank UGC, New Delhi for
financial support through a major research project sanctioned to VCK. VCK also wishes
to acknowledge Associateship of IUCAA, Pune, India.

\end{document}